\documentclass[11pt]{cernrep}

\newif\ifpreprint
\preprinttrue     % uncomment this line for preprint version

\usepackage{epsf}
\usepackage{graphicx}
\usepackage{here}

\ifpreprint
\usepackage{fancyhdr}
\fi

\begin{document}

\ifpreprint\pagestyle{empty}\fi

\ifpreprint
  \title{\rightline{\normalsize UASLP-IF-03-007}\huge Instrumentation}
\else
 \title{Instrumentation}
\fi
\author{J\"urgen Engelfried}
\institute{Instituto de F\'{\i}sica,
Universidad Aut\'onoma de San Luis Potos\'{\i}, Mexico
\ifpreprint \\ {\tt jurgen@ifisica.uaslp.mx, 
http://www.ifisica.uaslp.mx/\~\,jurgen/} \fi
}
\maketitle

\begin{abstract}
In this course, given at the school in 3 parts of 75~minutes each, we will
discuss the physics of particle detection, the basic designs and working
principles of detectors, and, as an example with more details, some
detectors for particle identification.
\end{abstract}
\ifpreprint
\footnotetext{Lecture course given at the 
         2nd Latin American School of High Energy Physics,
         San Miguel Regla, Mexico, June 1-14, 2003. To be published in
a CERN Yellow Report.}
\fi

\section{Introduction}

The detection and identification of particles and nuclei is not only
important in high-energy physics, but also in cosmic ray and nuclear physics.
The basic idea is that every effect of particles or radiation 
can be used as a working principle for a detector.
The main purpose is the 
detection and identification of particles with mass $m$ and charge $z$.
In particle physics, the charge is usually $z=0, \pm1$, but in nuclear, 
heavy ion physics, or cosmic rays, also higher charges are possible.

Examples of effects used for particle detection and identification 
include the momentum $p$ measured by deflecting
a charged particle (charge $z$) in a magnetic 
field ($\rho\propto{{p}/{z}}={{\gamma m\beta c}/{z}}$); 
the velocity $v$ measured by the time of flight $t$
($\beta={{v}/{c}}\propto {{1}/{t}}$) or the Cherenkov angle
($\theta_C={{1}/({\beta n}})$), the total energy $E$ measurements with 
a calorimeter, and for charge $z$ measurement the ionization energy loss
(${{dE}/{dx}}\propto z^2$).
With all the information together one can determine the
quadri-vector of the particle.

The basic detection techniques work mostly for charged particles only;
neutral particles are usually detected indirectly via reactions
producing charged particles.

Designing instrumentation and detectors requires knowledge of
the basic physics of interactions of charged and neutral particles
with matter,
mechanical engineering,
electrical engineering (high voltage),
electronic engineering,
interfaces to trigger, data acquisition and computing,
software engineering (calibration),
and operation (stability).
To know any one of them is not sufficient;
all have to be applied together to build, operate and use
an instrument for a physics measurement.
The final goal of the measurement and the precision (resolution) needed 
has always to be kept in mind, in order not to over-design the instrument.

The recommended literature includes, in addition to references throughout the
lecture, the following:
{\it Particle Detectors} by Claus Grupen~\cite{grupen},
{\it Detectors for particle radiation} by Konrad Kleinknecht~\cite{klein},
{\it Introduction to Experimental Particle Physics} by 
Richard C.~Fernow~\cite{fernow},
{\it Calorimetry} by Richard Wigmans~\cite{wigmans},
lecture notes and proceedings of ICFA Instrumentation Schools~\cite{icfa},
which are held bi-yearly since 1987, recently also as regional schools, 
and the {\it Particle Data Book}~\cite{pdg}, which contains short summaries
of important topics.  In the new 2004 edition~\cite{pdg04} a rewritten and
upgraded section
about instrumentation will be included. 

\newpage
\ifpreprint
\pagestyle{fancy}
\fancyfoot{ }
\fancyhead[R]{\small \thepage}
\fancyhead[L]{ }
\fancyhead[C]{ }
%\fancyhead{ }
%\fancyfoot[R]{\small \thepage}
\fancyfoot[L]{ }
\fancyfoot[C]{ }
\renewcommand{\headrulewidth}{0pt}
\fi

\leftline{\Large\bf Part I: Physics of Particle Detection}
\section{Interactions of Charged Particles}
\subsection{Kinematics}

To introduce variables used throughout this lecture, let us review some
basic kinematic formulas:

The maximum kinetic energy an electron (mass $m_e$) can gain when a 
particle with mass $m$, Energy $E$, momentum $p$, 
and velocity $v=\beta c$ collides with it is (by conservation of
momentum and energy)
\begin{equation}
E^{\rm max}_{\rm kin} = {{2 m_e c^2 \beta^2 \gamma^2}\over
{1+2\gamma{{m_e}\over{m}}+\left({{m_e}\over{m}}\right)^2}} =
{{2 m_e p^2}\over{m^2 + m_e^2 + 2m_e E/c^2}}
\end{equation}
In the limit of low energy
($2\gamma{{m_e}\over{m}}\ll1$)
and a massive particle ($m\gg m_e$), this leads to
\begin{equation}
E^{\rm max}_{\rm kin} = 2 m_e c^2 \beta^2 \gamma^2
\end{equation}
and in the relativistic limit ($E_{\rm kin}\approx E$, $pc\approx E$)
\begin{equation}
E^{\rm max} = {{E^2}\over{E+m^2c^2/2m_e}}
\end{equation}
In the special case of an electron-electron collision ($m=m_e$)
\begin{equation}
E^{\rm max}_{\rm kin} = {{p^2}\over{m_e+E/c^2}} =
{{E^2-m_e^2c^4}\over{E+m_e c^2}} = E - m_e c^2
\end{equation}

The scattering angle $\Theta$ of a particle with charge $z$ on a
nucleus with charge $Z$,
under an impact parameter $b$ is given by
\begin{equation}
\Theta = {{2z\cdot Z\cdot e^2}\over{\beta c b}} \cdot {{1}\over{p}}
\label{scatangle}
\end{equation}
and leads to the Rutherford cross section $\sigma$, given by
\begin{equation}
{{d\sigma}\over{d\Omega}} = {{z^2 Z^2 r_e^2}\over{4}}
\left({{m_ec}\over{\beta p}}\right)^2 {{1}\over{\sin^4\Theta/2}}
\end{equation}
For multiple scattering by a particle passing matter of
Radiation Length $X_0$ (see equation~\ref{radlen}) the root mean square
angle (the average angle is $0$) in a plane is given by
\begin{equation}
\sqrt{\langle\Theta^2\rangle} = \Theta_{\rm plane} =
{{13.6\,\mbox{MeV}}\over{\beta c p}}z\cdot\sqrt{{{x}\over{X_0}}}
\left\{ 1+0.038 \ln\left({{x}\over{X_0}}\right)\right\}
\label{multscat}
\end{equation}
and in space by
\begin{equation}
\Theta_{\rm space} = \sqrt{2}\,\Theta_{\rm plane}
\end{equation}
Usually this distribution is approximated by a 
Gaussian
\begin{equation}
P(\Theta)\,d\Theta = {{1}\over{\sqrt{2\pi}\Theta_{\rm plane}}} 
{\rm exp}\left\{-{{\Theta^2}\over{2\Theta_{\rm plane}^2}}\right\}\, d\Theta
\end{equation}
which describes well low angle behavior but underestimates the frequency
of large angle scatters.

\subsection{Energy Loss of Charged Particles}
The interaction of charges particles with matter is dominated by the
electromagnetic interaction, via the exchange of virtual or real photons.
Virtual photons are absorbed by the atoms of the material which leads
to ionization and/or  excitation of the atoms.
Real photons, also called radiation, is emitted by a charged particle if:
1.~its velocity is faster than the velocity of light in the medium
($v>c/n$) and is called {\sl  Cherenkov radiation}. 
2.~if the vector velocity relative to the phase velocity of
photons in matter ($\vec{v}/c_{\rm ph}=\vec{v}\cdot n/c$)
changes, which has different names. A change of 
$|\vec{v}|$ is called {\sl  Bremsstrahlung}, a change in the direction
of $\vec{v}$  {\sl Synchrotron radiation}
and a change in $n$ {\sl Transition Radiation}.

\subsubsection{Mean Ionization Energy Loss (Bethe-Bloch formula)}
We follow here the derivation in~\cite{grupen}.
The momentum transfer per target electron, for an impact parameter $b$
(see also equation~(\ref{scatangle})) is given by
\begin{equation}
p_b={{2r_em_ec}\over{b\beta}}z
\end{equation}
The Energy transfer $\epsilon$ in this collision is given in a
classical approximation by
\begin{equation}
\epsilon={{p_b^2}\over{2m_e}}={{2r_e^2 m_e c^2}\over{b^2\beta^2}}z^2
\label{etrans}
\end{equation}
The interaction rate $\Phi$, measured in $\mbox{cm}^2\mbox{/g}$, with an 
atomic cross section $\sigma$, is given by
\begin{equation}
\Phi[\mbox{g}^{-1}\mbox{cm}^2] =
{{N_A}\over{A}}\,\sigma[\mbox{cm}^2/\mbox{atom}]
\end{equation}
where $N_A$ is Avogadro's number, and $A$ atomic mass of the material.
In an annulus of area
$2\pi b\,db$, and $Z$ electrons per atom, the interaction rate is
\begin{equation}
\Phi(\epsilon)\,d\epsilon = {{N_A}\over{A}}2\pi b\,db\,Z
\end{equation}
equation~(\ref{etrans}) can be re-written to
\begin{equation}
b^2={{2 r_e^2 m_e c^2}\over{\beta^2}}\,z^2\,{{1}\over{\epsilon}}
\end{equation}
leading to an interaction rate as function of the energy transfer
\begin{equation}
\Phi(\epsilon)\,d\epsilon = {{N_A}\over{A}}\,\pi\,
{{2 r_e^2 m_e c^2}\over{\beta^2}}\,z^2\,Z\,{{d\epsilon}\over{\epsilon^2}}
\end{equation}
The energy loss $dE$ in a small path length $dx$ 
is simply the integral over all energy transfers
\begin{equation}
-{{dE}\over{dx}} = {{2\pi N_A}\over{A}}\, Z \int\limits_0^\infty\,\epsilon\,b\,db
=2\pi\,{{Z\,N_A}\over{A}}\,{{2r_e^2m_ec^2}\over{\beta^2}}\,z^2\,
\int\limits_0^\infty\,{{db}\over{b}}
\label{eqint}
\end{equation}
We are now left with the problem that this 
Integral is divergent for $b\to0$ and $b\to\infty$.

For $b\to0$ we assume a minimum impact parameter $b_{\rm min}$ given by 
half the de~Broglie wavelength of the particle; smaller structures can
not be resolved.
$b_{\rm min} = {{h}/({2p})} = {{h}/({2\gamma m_e \beta c})}$.
For $b\to\infty$, we make a model that for large $b$, the interaction
time $\tau_i$ will be larger than the 
revolution time $\tau_R$ of the electrons in the target atom.  The atom
will then appear neutral to the passing particle, and no interaction
will happen. The interaction time is the Lorentz-contracted fly-by time 
$\tau_i = b\sqrt{1-\beta^2}/v$, and the revolution time can be parametrized
by $\tau_R = {{1}/({v_Z\cdot Z})} = {{h}/{I}}$, $I$ being the
mean excitation energy of the atom, where
$I \approx 10\,\mbox{eV}\cdot Z$. Requesting
$\tau_i = \tau_R$ leads to $b_{\rm max} = {{\gamma h \beta c}/{I}}$.
Finally, integrating equation~(\ref{eqint}), we obtain
\begin{equation}
-{{dE}\over{dx}} = 2\pi{{Z\,N_A}\over{A}}{{2r_e^2m_ec^2}\over{\beta^2}} z^2
\big[ \ln{{2\gamma^2\beta^2m_ec^2}\over{I}} - \eta\big]
\end{equation}
The parameter $\eta$ takes into account the screening effect from nearby
atoms. The previous derivation assumed one single atom.

A more exact treatment, summarized in~\cite{pdg}, gives as the final result
for the Bethe-Bloch formula
\begin{equation}
-{{dE}\over{dx}} = 
2\pi{{Z\,N_A}\over{A}}{{2\,r_e^2\,m_e\,c^2}\over{\beta^2}} z^2
\Big[ {{1}\over{2}}\ln{{2\,m_e\,c^2\gamma^2\beta^2}\over{I^2}} 
\beta^2 - {{\delta}\over{2}}\Big]
\label{bethe-bloch}
\end{equation}
The density correction takes a more complicated form, and can be described by
\begin{equation}
{{\delta}\over{2}} = \ln {{\hbar\omega_p}\over{I}} + \ln\beta\gamma 
-{{1}\over{2}}\mbox{~~~~~}
\hbar\omega_p = \sqrt{4\pi N_e r_e^2}\, {{m_e\,c^2}\over{\alpha}}
\end{equation}
where
$N_e$ is the electron density and $\omega_p$ the
plasma frequency of the absorbing material,
and $\alpha$ the Sommerfeld fine structure constant.

\begin{figure}
\begin{center}
\leavevmode
\epsfxsize=8cm
\epsffile{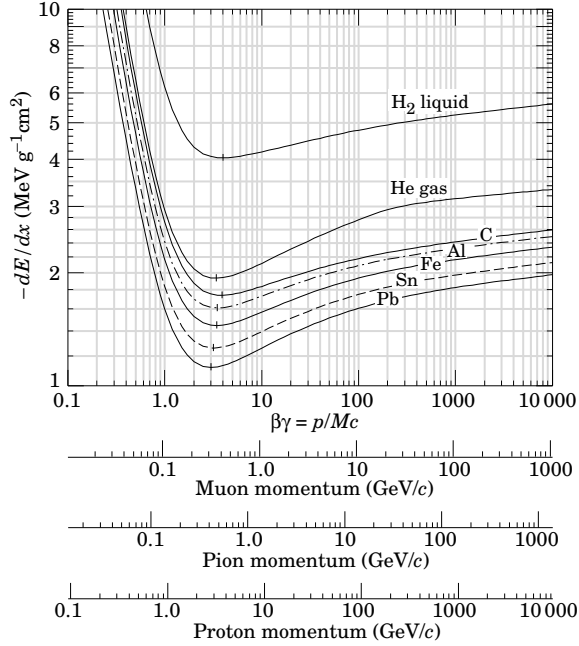}
\caption{Mean energy loss for different materials~\cite{pdg}.}
\label{pdg263}
\end{center}
\end{figure}

In fig.~\ref{pdg263} (taken from~\cite{pdg}) we show examples for energy
loss in different materials.
The minimum energy loss of all particles in nearly all materials occurs
at $3\le\beta\gamma\le4$, and is in the range of $\mbox{MeV/(g/cm}^2)$
(examples: helium: $-dE/dx = 1.94\,\mbox{MeV/(g/cm}^2)$,
uranium: $1.08\,\mbox{MeV/(g/cm}^2)$)
with the exception of hydrogen, in which particles experience
a larger energy loss ($Z/A=1$).
Due to the $\ln\gamma$ term the energy loss increases for relativistic
particles and reaches the so-called 
Fermi-Plateau due to the density effect.
In  gases the plateau is typically 
$\approx60\,\mbox{\%}$ higher as the minimum.

Due to the increased energy loss at smaller $\beta\gamma$, particles
will deposit most of their energy at the end of their track, just before
they will be completely stopped. This ``Bragg Peak'' is used for the
treatment of deep-seated tumors, selecting the particle type and
initial energy to optimize the energy loss close to the tumor location,
avoiding too much damage to tissue above the tumor.

\subsubsection{Landau Distribution of the energy loss}
The energy loss is distributed asymmetrically around the mean
energy loss described by the 
Bethe-Bloch formula (eq.~\ref{bethe-bloch}); the distribution
can be approximated by the Landau distribution $\Omega(\lambda)$
\begin{equation}
\Omega(\lambda)={{1}\over{\sqrt{2\pi}}}\,
e^{-{{1}\over{2}}(\lambda+e^{-\lambda})}
\label{landau}
\end{equation}
where $\lambda = {{({{dE}\over{dx}}) - 
({{dE}\over{dx}})^{\rm m.p.}}/({0.123\,\mbox{keV}})}$, with
$({{dE}\over{dx}})^{\rm m.p.}$ being the most probable energy loss.
In gases and thin absorbers the Landau fluctuations have to be considered;
for example
a particle with $\beta\gamma=4$  in Argon experience a most probable energy
loss of 
$({{dE}\over{dx}})^{\rm m.p.}=1.2\,\mbox{keV/cm}$ and a mean energy loss of
$\langle {{dE}\over{dx}}\rangle=2.69\,\mbox{keV/cm}$.

For particle identification, one has to sample often (typically 160 times or
more) to measure the Landau distribution and use adequate algorithms 
(e.g.\ ``truncated mean'') to obtain the mean energy loss.

Examples for successful application of this particle identification
method can be found in the literature~\cite{peptpc,jade,opal}.

\subsection{Scintillation}
\label{scint}
Scintillation is the de-excitation of previously exited atoms, molecules
or materials. It appears in inorganic crystals, organic liquids or plastics,
and gases.

In inorganic crystals, scintillation is usually an effect of the lattice.
Energy lost by a passing particle is used to produce electron-hole pairs
or excitons, which de-excite at activation centers (crystal imperfections)
emitting photons.  Common materials used include Thallium-doped NaI and CsI,
BaF$_2$, and BGO (Bismuth-Germanium-Oxide).

Organic scintillators in liquids or plastic come in three components:
the primary scintillator (antracene C$_{14}$H$_{10}$, 
naphthalene C$_{10}$H$_{8}$); a wavelength shifter 
(POPOP C$_{24}$H$_{16}$N$_2$O$_2$ 1.4-Bis-[2-(5-phe\-nyl\-oxazolyl)]-benzene,
BBO C$_{27}$H$_{19}$NO 2.5-di-(4-biphenyl)-oxazole)
and a base material like mineral oil or
PMMA (C$_5$H$_8$O$_2$ polymethylmetacralate).

In gases, the atoms or molecules previously excited or ionized 
by $dE/dx$ emit photons due to recombination or de-excitation. This
is often a background and noise source when using effects for measurements, 
for example Cherenkov.
Basically all gases (Xe, Kr, Ar, N$_2$) exhibit this behavior.

The light yield (number of photons $N$) 
is a non-linear function of the energy and shows a saturation
behavior, usually described by Birk's formula
\begin{equation}
N=N_0\, {{dE/dx}\over{1+k_B\cdot(dE/dx)}}
\end{equation}
where $k_B\approx 0.01\,\mbox{g/MeV}\,\mbox{cm}^2$ is the
so-called Birk's density Parameter. 
Recently some better approximation were developed by a 
Mexican group~\cite{menchaca}.

Typically about $100\,\mbox{eV}$ energy loss are necessary to 
produce one photon.
The addition of wavelength shifters is necessary to avoid self-absorption
of the produced photons.

\subsection{Cherenkov Radiation}
\label{cheren}
 A charged particle with a velocity $v$ larger than the velocity of light
 in a medium emits light~\cite{cherenkov}.
The threshold velocity $v_{\rm thres}$ is given by
\begin{equation}
\beta_{\rm thres} = {{v_{\rm thres}}\over{c}}\ge {{1}\over{n}},
\mbox{~~or~~~}
\gamma_{\rm thres} = {{n}\over{\sqrt{n^2-1}}}
\label{cherenkovthres}
\end{equation}
$n$ being the 
(wavelength dependent) refractive index of the material.
The angle of emission is given by
\begin{equation}
\cos\theta_c = {{1}\over{\beta\, n}} = {{1}\over{{{v}\over{c}}\, n}}
\label{jurgen:cherenkovangle}
\end{equation}
with a maximum angle of $\theta_c^{\rm max} = \arccos{{1}/{n}}$.
For water $\theta_c^{\rm max} = 42^\circ$, for
Neon at $1\,\mbox{atm}$ $\theta_c^{\rm max} = 11\,\mbox{mrad}$
The number of photons emitted by Cherenkov radiation is~\cite{francktamm}
\begin{eqnarray}
{{d^2N}\over{dE dl}} &=& {{\alpha z^2}\over{\hbar c}}
\left(1 - {{1}\over{(\beta n)^2}}\right)
= {{\alpha z^2}\over{\hbar c}} \sin^2\theta_c\\
{{d^2N}\over{d\lambda dl}}
&=& {{2 \pi \alpha z^2}\over{\lambda^2}} \sin^2\theta_c
\label{jurgen:dndl}
\end{eqnarray}

Water Cherenkov counters were originally developed to set limits
on the lifetime of protons, but got converted for (solar) neutrino
detection. On example is (Super-)Kamiokande.  Cherenkov effect in
water is also used in the tanks of the Auger Experiment to detect
muons produced in cosmic ray air showers.

The simplest Cherenkov devices are threshold Cherenkov counter,
only using the difference between no light or light to set ranges
for the velocity of particles. A recent example is the
Belle Detector~\cite{belle}.

Ring Imaging Cherenkov Detectors not only use the threshold information,
but actually measure with the help of  some imaging system (see also
section~\ref{rich}) the Cherenkov angle, obtaining the velocity of 
the particle.  A recent example for this is the SELEX RICH~\cite{selexrich},
were we show the dependency of the ring radius (or the Cherenkov angle) on
the particle momenta.
\begin{figure}
\epsfxsize=\hsize
\epsffile{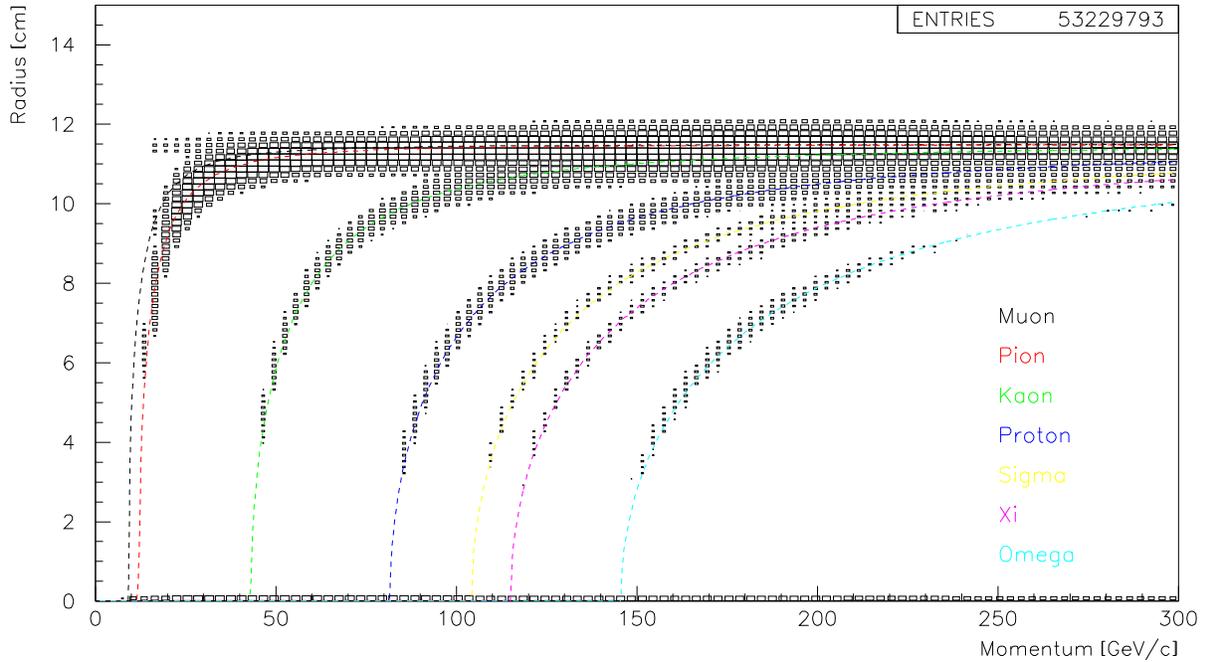}
\caption{Radius of imaged Cherenkov photons measured by the
SELEX RICH~\cite{selexrich} as function of the momentum for 53 million
single negative track events}
\end{figure}

\subsection{Transition Radiation}
\label{tr}
Transition radiation is emitted due to the
reformation of the particle field while traveling from
a medium with dielectric constant
$\epsilon_1$ to a medium with $\epsilon_2$.
The energy of the radiation emitted at a single interface $S$ is given by
\begin{equation}
S = {{\alpha\hbar z^2}\over{3}}
              {{(\omega_1 - \omega_2)^2}\over{\omega_1 + \omega_2}}\gamma
\label{transrad}
\end{equation}
were $\omega_1$ and $\omega_2$ are the plasma frequencies.
Typical values are for air $\omega_1=0.7\,\mbox{eV}$ and polypropylene 
$\omega_2=20\,\mbox{eV}$.
The spectral and angular dependence of Transition Radiation is 
\begin{equation}
{{d^2}\over{d\vartheta d\omega}} = {{2e^2}\over{\pi c}}
\left( {{\vartheta}\over{\gamma^{-2} + \vartheta^2 + \omega_1^2/\omega^2}} -
 {{\vartheta}\over{\gamma^{-2} + \vartheta^2 + \omega_2^2/\omega^2}}\right)^2
\end{equation}
Most of the radiation is emitted in a cone with half angle $1/\gamma$, 
forward in the particle direction.
For large photon energies $\omega>\gamma\omega_2 \approx 20-30\,\mbox{KeV}$,
the intensity drops like $\propto \gamma^4/\omega^4$,
at medium energies $\gamma\omega_1\lappeq\omega\lappeq\gamma\omega_2$,
we can observe a
logarithmic decrease with $\omega$, and at
small energies $\omega<\gamma\omega_1\approx 1\,\mbox{KeV}$,
the intensity is almost constant.
The probability to emit a $\mbox{KeV}$ photon is only $\approx 10^{-2}$;
to  have a detectable rate, one needs to use a lot of interfaces,
usually a stack of radiator foils.

A minimum foil thickness is needed for the particle field to reach an
equilibrium inside the medium.
The transitions into and out of the foil produce the same radiation 
($(\omega_1-\omega_2)^2$ in eq.~\ref{transrad}), so 
interference effects can occur, also when the foils are space equally.
Part of the radiation is re-absorbed in subsequent foils
($\propto Z^5$), and low-$Z$ materials 
(mylar, CH$_2$, carbon fibers, lithium,
thickness $30\,\mu\mbox{m}$, distance: $300\,\mu\mbox{m}$ 
are consequently used.

\subsection{Bremsstrahlung}
\label{bremsstrahlung}
Bremsstrahlung is emitted when a charged particle experiences some
acceleration.  The change in energy is given by
\begin{equation}
-{{dE}\over{dx}} = 4\alpha N_A\, {{Z^2}\over{A}}\,z^2\,
\left({{e^2}\over{mc^2}}\right)^2\,E\,\ln{{183}\over{Z^{1/3}}}
\label{brems}
\end{equation}
which can be used to define via $-{{dE}\over{dx}} = {{E}\over{X_0}}$
the Radiation Length $X_0$.
The PDG~\cite{pdg} gives a fit to estimate $X_0$ for different materials:
\begin{equation}
X_0={{716.4\,A}\over{Z(Z+1)\,\ln(287/\sqrt{Z})}}\, [\mbox{g/cm}^2]
\label{radlen}
\end{equation}
Typical values for $X_0$ are given in table~\ref{x0}.
The particle energy where the ionization energy loss is
equal to the energy loss by Bremsstrahlung is called the {\sl Critical Energy}
and is given for electrons by
$E^e_{\rm crit} = {{610\,{\rm MeV}}/({Z+1.24})}$ for solids and liquids
and
$E^e_{\rm crit} = {{710\,{\rm MeV}}/({Z+0.92})}$ for gases.
\begin{table}
\begin{center}
\begin{tabular}{|l|c|c|c|}
\hline
\multicolumn{1}{|c|}{Material~~~~~}&$X_0\,[\mbox{g/cm}^2]$~~~~~&
$X_0\,[\mbox{cm}]$~~~~~&
$E^e_{\rm crit}\,[\mbox{MeV}]$~~~~~\\
\hline
~~~air&37&30000&84\\
~~~iron&13.9&1.76&22\\
~~~lead&6.4&0.56&7.3\\
\hline
\end{tabular}
\end{center}
\caption{Typical values for the Radiation length $X_0$}
\label{x0}
\end{table}
As seen in eq.~\ref{brems}, $-dE/dx\propto 1/m^2$, so Bremsstrahlung from
electrons dominates; but for muons in iron $E^\mu_{\rm crit} = 960\,\mbox{GeV}$
so at TeV energies we have to rethink the ``all penetrating $\mu$''
and can consider Muon Calorimetry.
Bremsstrahlung is an important component of electromagnetic cascades
(Calorimetry), which will be discussed in section~\ref{emshower}

\subsection{Direct Electron Pair Production}
In the Coulomb field of a nucleus electron pairs can be produced
directly via virtual photons. The corresponding energy loss is 
$\propto E$ for large Energies, the same behavior as in the
case of Bremsstrahlung and nuclear interactions (see sections~\ref{nuclint}).
The range of muons in rock is heavily influenced by this interaction,
and yields
$140\,\mbox{m}$ ($E_\mu=100\,\mbox{GeV}$),
$800\,\mbox{m}$ ($E_\mu=1\,\mbox{TeV}$), and
$2300\,\mbox{m}$ ($E_\mu=10\,\mbox{TeV}$).  For this reason experiments 
requiring low backgrounds have to be performed in deep (gold-)mines to be
shielded from atmospheric muons produced by cosmic rays. 

\section{Nuclear Interactions}
\label{nuclint}
This effect is important for the detection of neutral particles.
The total cross section 
$\sigma_{\rm tot}\approx 50\,\mbox{mbarn}$ can be separated in an elastic
and an inelastic part; the latter is $\sigma_{\rm inel} \propto A^\alpha$,
with $\alpha=0.71$.  As in the Bremsstrahlung case (eq.~\ref{radlen}), a
{\sl Nuclear Interaction length} $\lambda_I$ and a
{\sl Nuclear Absorption length} $\lambda_a$ can be defined via
$\lambda_I = A/(N_A\,\rho\,\sigma_{\rm tot})$ and
$\lambda_a = A/(N_A\,\rho\,\sigma_{\rm inel})$.
Typical values are shown in table~\ref{lambdaI}.
\begin{table}
\begin{center}
\begin{tabular}{|l|c|c|c|c|}
\hline
&Al&Fe&Pb&Air\\
\hline
$\lambda_I\, [\mbox{cm}]$&26.2&10.6&10.4&48000\\
$\lambda_I\, [\mbox{g/cm}]$&70.6&82.8&116.2&62.0\\
\hline
\end{tabular}
\caption{Nuclear Interaction length $\lambda_I$ for some materials.}
\label{lambdaI}
\end{center}
\end{table}
By comparing with table~\ref{x0} one can conclude that for
most material $\lambda_I, \lambda_a > X_0$.

The multiplicity in nuclear inelastic interactions grows 
logarithmically with $E$, and the secondary particles leave with an
average transverse momentum $p_T\approx350\,\mbox{MeV}/c$.

\section{Interactions of Neutral Particles}

\subsection{Interactions of Photons}
Photons are, in contrary to charged particles, attenuated in matter;
The intensity~$I$ after a distance~$x$ is given by
$I = I_0\, e^{-\mu x}$, where $\mu$ is the so-called
{\sl Mass Attenuation Coefficient}.  Three basic processes contribute
to the $\mu = {{N_A}\over{A}}\sum_{i=1}^3 \sigma_i$: 
1)~The Photoelectric Effect;
2)~Compton Scattering;
3)~Pair Production.
Each of the processes has its own energy dependence, and we will discuss
them in the following sections.

\subsubsection{Photoelectric Effect}
\label{photoeffect}
One photon of sufficient energy is absorbed by an atom,
releasing one electron and leaving behind a positive ion.
The released electron comes predominantly from the  K-shell.
The effect presents a complicated energy and Z dependence,
\begin{equation}
\sigma_{\rm photo}^K = \left({{32}\over{\epsilon^7}}\right)^{1/2}\,
\alpha^4\, Z^5\,\sigma_{\rm Thomson} \mbox{~~}[\mbox{cm}^2/\mbox{atom}]
\end{equation}
with $\epsilon = {{E_\gamma}/{m_ec^2}}$ and
$\sigma_{\rm Thomson}=(8/3)\pi\,r_e^2=665\,\mbox{mbarn}$.
%In the limit of high photon energies,
%$\sigma_{\rm photo}^K = 4\pi\,r_e^2\,Z^5\,\alpha^4\,/{\epsilon}$

\subsubsection{Compton Scattering}
The Compton effect describes the elastic scattering of photon on a
(quasi-)free electron, 
a text book example to prove that the photon can be treated as a particle.
At high energies, the Compton cross section
$\sigma_C \propto {{\ln\epsilon}\over{\epsilon}}\cdot Z$
and the energy of the photon after scattering $E_\gamma^\prime$
is related to the scattering angle $\theta$ via
\begin{equation}
{{E_\gamma^\prime}\over{E_\gamma}} = 
{{1}\over{1+\epsilon(1-\cos\theta_\gamma)}}
\end{equation}
The maximum kinetic energy the electron can gain is
$E_{\rm kin}^{\rm max}(\theta_\gamma=\pi) = 
{{2\epsilon^2}\over{1+2\epsilon}}\,m_ec^2$
which goes in the limit of $\epsilon\gg1$ to 
$E_{\rm kin}^{\rm max}(\theta_\gamma=\pi) \to E_\gamma$.

\subsubsection{Pair Production}
\label{pairprod}
In the electric field of a nucleus (another text book example) a
photon with sufficient energy can convert into an electron-positron pair.
The threshold energy is given by
$E_\gamma \ge 2\,m_ec^2 + 2\,m_ec^2/m_{\rm target}$, leading to
$E_\gamma \ge \approx 2\,m_ec^2$ for pair production on a nucleus, and
$E_\gamma \ge 4\,m_ec^2$ for pair production on an electron.

The cross section for $E_\gamma \gg 20\,\mbox{MeV}$ is given by
\begin{equation}
\sigma_{\rm pair} = 4\,\alpha\,r_e^2\,Z^2\,
\left({{7}\over{9}}\ln{{183}\over{Z^{1/3}}} - {{1}\over{54}}\right)
\mbox{~~}[\mbox{cm}^2/\mbox{atom}]
\approx {{7}\over{9}}\,{{A}\over{N_A}}\,{{1}\over{X_0}}
\label{pair}
\end{equation}
with the definition for $X_0$ from eq.~\ref{radlen}.

\subsection{Interaction of Neutrons}
Neutrons are neutral particles; for their detection we have to
use some indirect methods. Neutrons have to interact and the
interaction has to produce charged particles, which can themselves be detected
with previously discussed techniques.

For low neutron energies ($<20\,\mbox{MeV}$), interactions like
$n+^6\!{\rm Li}\to \alpha + ^3\!{\rm H}$ using LiI(Tl) scintillators,
$n+^{10}\!{\rm B} \to \alpha + ^7\!{\rm Li}$ using BF$_3$ filled gas counters,
$n+^3\!{\rm He} \to p + ^3\!{\rm H}$ using 
$^3{\rm He}$-filled proportional counters,
and $n+p\to n+p$ using proportional chambers filled for example
with for example\ CH$_4$, are used.

For high neutron energies ($>1\,\mbox{GeV}$),
coated proportional counters to detect the fission products of 
$n+^{235}\!{\rm U}$ and hadron calorimeters (see section~\ref{hadcal}) 
are used.
 
\subsection{Interactions of Neutrinos}
Neutrinos and anti-neutrinos are directly detected with charged-current
interactions:
$\nu_e+n \to p+e^-$, $\overline{\nu_e}+p \to n+e^+$,
$\nu_\mu+n \to p+\mu^-$, $\overline{\nu_\mu}+p \to n+\mu^+$,
$\nu_\tau+n \to p+\tau^-$, $\overline{\nu_\tau}+p \to n+\tau^+$.
The cross section is extremely small, for neutrinos of $0.5\,\mbox{MeV/c}$
$\sigma(\nu_eN) = 1.6\cdot10^{-44}\,\mbox{cm}^2$
For high energies (GeV range), the cross section is $\propto E_\nu$,
$\sigma(\nu_\mu N) = 0.67\cdot10^{-38}\, E_\nu\,\mbox{cm}^2/\mbox{(nucleon GeV)}$ and
$\sigma(\overline{\nu_\mu} N) = 0.34\cdot10^{-38}\, E_\nu\,\mbox{cm}^2/\mbox{(nucleon GeV)}$.

Usually indirect measurements of the neutrinos are performed with hermetic
detectors, applying missing momentum and missing energy techniques.

\section{Electromagnetic Cascades}
\label{emshower}
As mentioned in section~\ref{bremsstrahlung}, for high electron energies the
most probable process is Bremsstrahlung (eq.~\ref{brems}), e.g.\ the
generation of a real photon.  For high photon energies, the most
probable process is pair creation (section~\ref{pairprod}, eq.~\ref{pair}).
We can understand the basic features of the two processes playing ping-pong
with a simple toy model. In every step~$t$,
which corresponds to one conversion,
where the step size is related to the Radiation length 
$X_0$ (eq.~\ref{radlen}),
the number of particles~$N$ is doubled ($N(t)=2^t$)
and the energy~$E$ of each particle is halved ($E(t) = E_0 \cdot 2^{-t}$).
The multiplication stops at $t_{\rm max}$, where the energy of 
the particle falls below
a critical energy $E_c=E_0\cdot2^{-t_{\rm max}}$, which we can 
associate to the pair production threshold.
For the position of the shower maximum we obtain
\begin{equation}
t_{\rm max} = {{\ln E_0/E_c}\over{\ln 2}} \propto \ln E_0
\end{equation}
For the total number of shower particles $S$:
\begin{equation}
S=\sum\limits_{t=0}^{t_{\rm max}} N(t) = \sum 2^t=2^{t_{\rm max}+1} - 1
\approx 2\cdot2^{t_{\rm max}} = 2\cdot {{E_0}\over{E_e}} \propto E_0
\end{equation}
and for the total track length $S^\star$, measured in sampling steps $t$
\begin{equation}
S^\star = {{S}\over{t}} = 2\cdot{{E_0}\over{E_c}} \cdot {{1}\over{t}}
\end{equation}
The energy Resolution is then given by
\begin{equation}
{{\sigma(E_0)}\over{E_0}} = {{\sqrt{S^\star}}\over{S^\star}} = 
{{\sqrt{t}}\over{\sqrt{2E_0/E_c}}} \propto {{\sqrt{t}\over{\sqrt{E_0}}}}
\end{equation}
All basic features of electromagnetic showers are reproduced by this simple
model: the position of the shower maximum is $\propto\ln E_0$, the number of
shower particles is $\propto E_0$, and the energy resolution is 
$\propto 1/\sqrt{E_0}$.

A more realistic description of longitudinal shower development yields
\begin{equation}
{{dE}\over{dt}} = {\rm const}\cdot t^a \cdot e^{-bt}
\end{equation}
where $a$, $b$ a parameters which have to be determined for the specific
detector material used.

The lateral spread of the shower, which is caused by multiple scattering,
is given by the Moli\`ere Radius $R_m$,
\begin{equation}
R_m = {{21\,\mbox{MeV}}\over{E_c}}\,X_0 [\mbox{g/cm}^2]
\label{moliere}
\end{equation}
which is defined so that
$95\,\mbox{\%}$ of the shower energy is contained in a 
cylinder of radius $2\,R_m$.
For a homogeneous calorimeters build out of iron $R_m^{\rm Fe}=1.8\,\mbox{cm}$,
and out of lead $R_m^{\rm Pb}=1.6\,\mbox{cm}$.

\section{Hadron Showers}
\label{hadronshower}
The longitudinal development of hadron showers is governed by the
nuclear interaction length $\lambda_I$ (see section~\ref{nuclint}), and
the lateral development by the transverse momentum $p_T$ of 
secondary particles.
Since $\lambda_I > X_0$ and the average $p_T$ from nuclear interactions
is much larger than the average $p_T$ from multiple
scattering (eq.~\ref{multscat}),
hadron showers 
are wider and longer than electromagnetic showers of the same original
energy.

Due to the multitude of possible processes during the shower creation,
the hadron energy is transferred into several groups, each of them with
their own systematic energy dependencies, and the distribution into the
groups is subject to fluctuations.
Part of the energy goes into charged particles, and if they are $\mu$'s,
some of the energy is lost (or better: not detected). $\pi^0$'s 
produced in the shower initiate an electromagnetic cascade, which is
usually contained.  Nuclear binding energy can be partially recovered
by fission, but the energy in 
nuclear fragments will be partially lost.
In general, the visible energy is systematically 
lower than the original particle energy,
and due to the fluctuations in 
the losses the energy resolution of hadron calorimeters is 
worse than for electromagnetic calorimeters.

One important point to optimize the energy resolution 
is to balance the different responses to electrons
and hadrons ($e/\pi = 1$, compensation).  We will discuss this again in the
section~\ref{hadcal}.

\section{Drift and Diffusion in Gases}
\label{drift}
In this section we will discuss effects after particle lost it's energy
and ionized the atoms or molecules of a gas.

The electrons and ions lose their energy by multiple elastic and
inelastic collisions with
the other atoms and molecules of the gas, and will shortly reach thermal
equilibrium.
The ionization cloud will diffuse and has a Gaussian width 
$\propto \sqrt(t)$, where $t$ is here the time.

With an additional electric field, the cloud will drift with a constant
velocity $v_{\rm drift}$, which depends on the gas and the strength of
the electrical field, but is usually measured in $\mbox{cm}/\mu\mbox{sec}$
(example: argon-isobutane: 
$v_{\rm drift}^{\rm electron} \approx 5\,\mbox{cm}/\mu\mbox{sec}$).
The longitudinal and transverse diffusion coefficients are
different, due to the energy dependent electron-molecule cross section.
The drift velocity is $v_{\rm drift} \propto 1/m$, so ions drift
several thousand times slower than electrons.

With an additional magnetic field, the drift will be under an angle
due to the additional (velocity dependent) Lorentz force.

\newpage
\leftline{\Large\bf Part II: Basic Design of Detectors}
\vspace{0.3cm}
In this part we will apply some of the physics discussed in the first part to
examples of particle detectors.  We will discuss
wire chambers in several incarnations, which are usually used
to measure the position of particles,
and electromagnetic and hadron calorimeters, 
which are used (as already indicated by their name) 
to measure the total energy.
In the third part we discuss specific detectors used to identify particles.

It is important to note that in general it is not possible to 
classify one detector type into one specific group; overlaps are common 
and useful so one detector can perform more than one specific task.

\section{Wire Chambers}

\subsection{Ionization Chamber}
This is the simplest gas detector, but also can be build with liquid or
solid materials. It is basically a large capacitor, either flat or with
cylindrical symmetry (a can with a wire in the center), 
where the ionization (electrons and ions)
drift to anode and cathode resp.  The moving charges can be collected,
amplified and integrated over time, which will give a 
signal which is proportional to the original ionization loss.

\subsection{Proportional Counters}
This detector is usually build as a cylinder like an 
ionization chamber, but with a smaller
diameter wire in the center or operated at a higher voltage (or both).
Due to the higher field close to the surface of the wire, the electrons
gain enough energy so they are able to ionize more of the counting gas
molecules and produce an avalanche.
This charge multiplication results in a higher induced current
in the wire.  The word {\sl proportional} in the name 
stems from the fact that at not
too high operating voltages the
gas amplification is constant and the measured signal is
proportional to the primary ionization.

\subsection{Geiger-M\"uller Counter}
This detector is build in the same geometry as the Proportional Counter, 
but operated at an even higher voltage. 
In the gas multiplication avalanche, in addition to ions also excited
atoms and molecules are produced, where some of them de-excite via the emission
of a photon.  This leads to a copious production of photons, which can
ionize more via the photoelectric effect (see section~\ref{photoeffect}),
also far away from the original avalanche.

Two possibilities are used to stop discharge.  One is to make the
resistor between the high voltage supply and the counter big enough, so every
time the counter draws a lot of current, the voltage will drop.
The other is to add
alcohols, like methylal or ethyl-alcohol, or hydrocarbons, like methane,
ethane, or isobutane, to the counting gas (usually argon); they
will absorb the UV photons and reduce the free path for them.

\subsection{Gas Multiplication}
As seen in the previous sections, the same detector can be operated in
different modes, and each of them carries its own name.  At very low voltages
(or electric fields), the ions and electrons will just recombine. At low
voltages, the two charges are separated and we operate a ionization chamber.
For medium voltage, we have constant gas amplification and are operating
a proportional counter, and a high voltages, we operate in the Geiger-Mode.
At very high voltage we will experience glow discharge, which can
destroy the detector.

\subsection{Multiwire Proportional Chambers}
\label{mwpc}
At the beginning of the 1960's, George Charpak (Nobel Price 1992) and
others started to build detectors containing more than one counting wire,
extending significantly the use of this kind of detectors.  The advantage
to the previously used bubble chambers is obvious:  The signal is electric,
so it can can easily be fed into electronics, which also started to be
developed around that time.

\begin{figure}[h]
\includegraphics[width=0.49\textwidth]{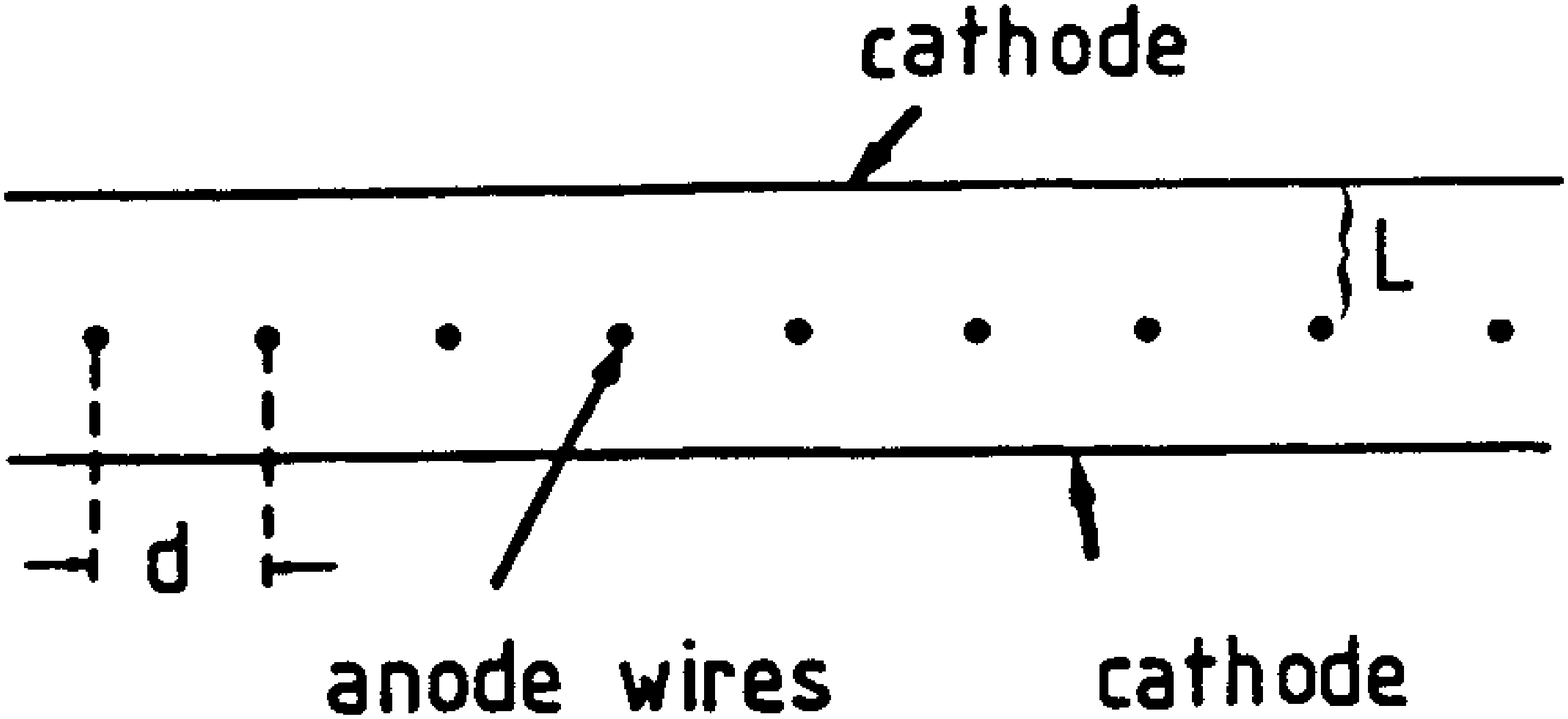}
\hfill
\includegraphics[width=0.49\textwidth]{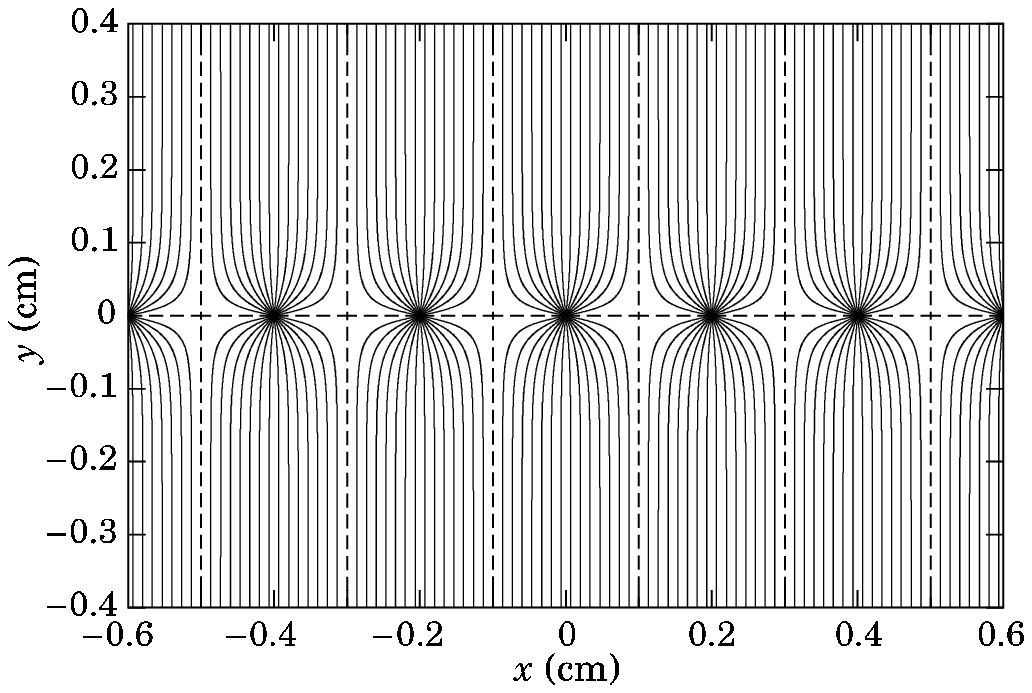}
\caption{Left: Side view of a Multiwire Proportional Chamber.
Right: Electric field lines in a MWPC.}
\end{figure}
Over a (usually rectangular) frame, several wires are stretch in
parallel.  A cathode foil above and below the wire plane completes the
detector.  Typical dimensions are: wire distance $d$ a few mm
(smallest known to me $0.8\,\mbox{mm}$), wire diameter
$10\,\mu\mbox{m} - 30\,\mu\mbox{m}$, thickness of chamber $L$ a few wire
distances, and total sizes up to square meters.
The total size is limited by the wire tension and electrostatic repulsion.

The electric field lines form automatically cells, the electrons
from ionization drift to the closest wire. The readout electronics
consists of a simple discriminator sending a logical output whenever
a particle passed close to that wire.
The spatial resolution is  $\sigma = d/\sqrt{12}$, but only 
in one dimension.  A second, identical module, mounted under an angle,
has to be used for the other dimension.  Some detectors also use
segmented cathodes for that.

\subsection{Planar Drift Chambers}
\label{driftchamber}
As mentioned above, wire chambers are limited by the
electrostatic repulsion between wires, which requires a minimum distance
as well as strong frames to stand the forces. At the middle of the 1960's, 
Heintze and Walenta~\cite{heintze}, 
but also others, came up with a new idea:  Keep
the basic geometry of a MWPC, but make the wires further apart, and use
the drift time of the electrons to the wire as additional information.
As described in section~\ref{drift}, in an homogeneous electric field
the drift velocity is constant.  A start signal for measuring the drift 
time is usually readily available in an experiment, for example from
a plastic scintillator (section~\ref{scint}) defining the incoming beam.

Drift distances over several centimeters can be easily achieved, and the
resolution is limited by diffusion;  In small chambers,
$\sigma>20\,\mu\mbox{m}$, in larger (square meter)
chambers a few $100\,\mu\mbox{m}$  can be obtained.

\subsection{Cylindrical Drift Chambers / Jet Chambers}
The same idea as in the planar drift chambers can be realized in
a cylindrical geometry, more adapted to collider detectors.  The wire
will be (nearly) parallel to the beam pipe, and a magnet field parallel
to the beam (so it will not disturb it) will curve the
charged particles in a plane perpendicular to the beam direction,
allowing the determination of the momentum and charge of the particles.
Wires are arranged radially to the outside, and the detector looks like
composed of pieces of pie in the $r$-$\phi$ view.  The largest chamber
of this type is the Central Jet Chamber of the OPAL detector at LEP,
with a radius of nearly 2~meters, and a length of 4~meters, containing
about 3500~counting wires~\cite{opal}.

\subsection{Time Projection Chamber}
TPCs are another device which uses in a clever way drifting electrons
in a gas counter.  A big usually cylindrical volume (meters in diameter, 
meters in length) has a (thin) positive
high voltage electrode in the center plane, forcing electrons to drift
toward the end plates of the cylinder.  To reduce the longitudinal
diffusion of the electrons, a magnet field is in parallel with the
electrical field, so electrons can only spiral around the magnet field
lines.  The end plates are equipped with wires and a segmented cathode plan
taking a two-dimensional image of the arriving electrons; the third coordinate
is obtained by measuring the drift time.

This detector was successfully used in the ALEPH experiment at LEP, but
also is used in Heavy Ion experiments with extremely high track density,
like in STAR at RHIC and, in preparation, for ALICE at LHC.

\section{Solid State Tracking Detectors}

\subsection{Silicon Strip Detectors}

In the 1980's, silicon microstrip detectors became used heavily in HEP.
They are absolutely necessary to measure properties of particles
containing charm and beauty quarks.  Examples for very successful experiments
using this kind of detectors include E691 at Fermilab, WA82 at CERN,
and, in colliders, CDF, the 4 LEP experiments (Aleph, DELPHI, L3, OPAL), and
the HERA experiments.  Today there are a lot of experiments using
silicon microstrips, with channel counts up to 1~million or more.

\begin{figure}[htb]
\begin{center}
\leavevmode
\vspace{6cm}
\includegraphics{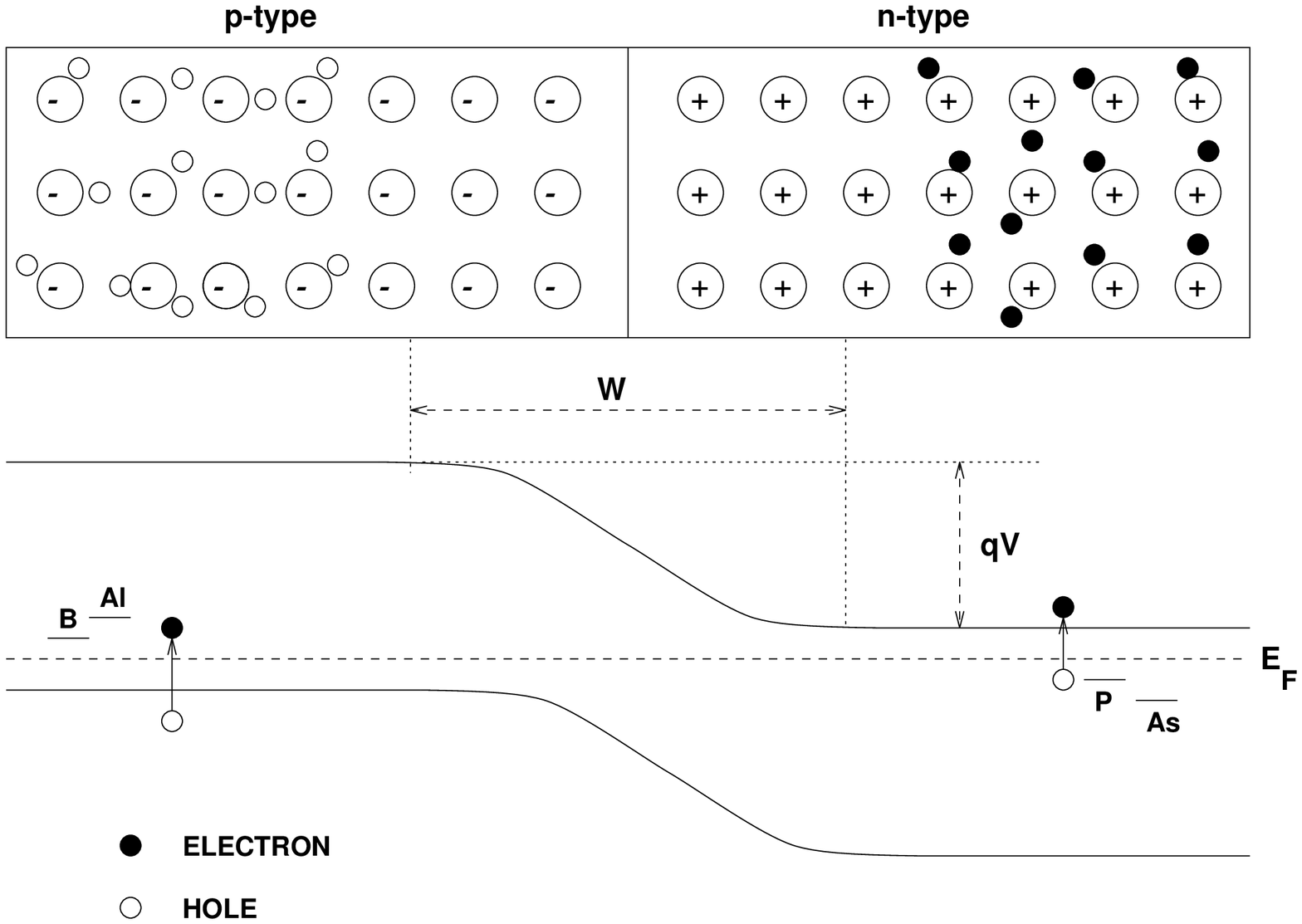}
\includegraphics{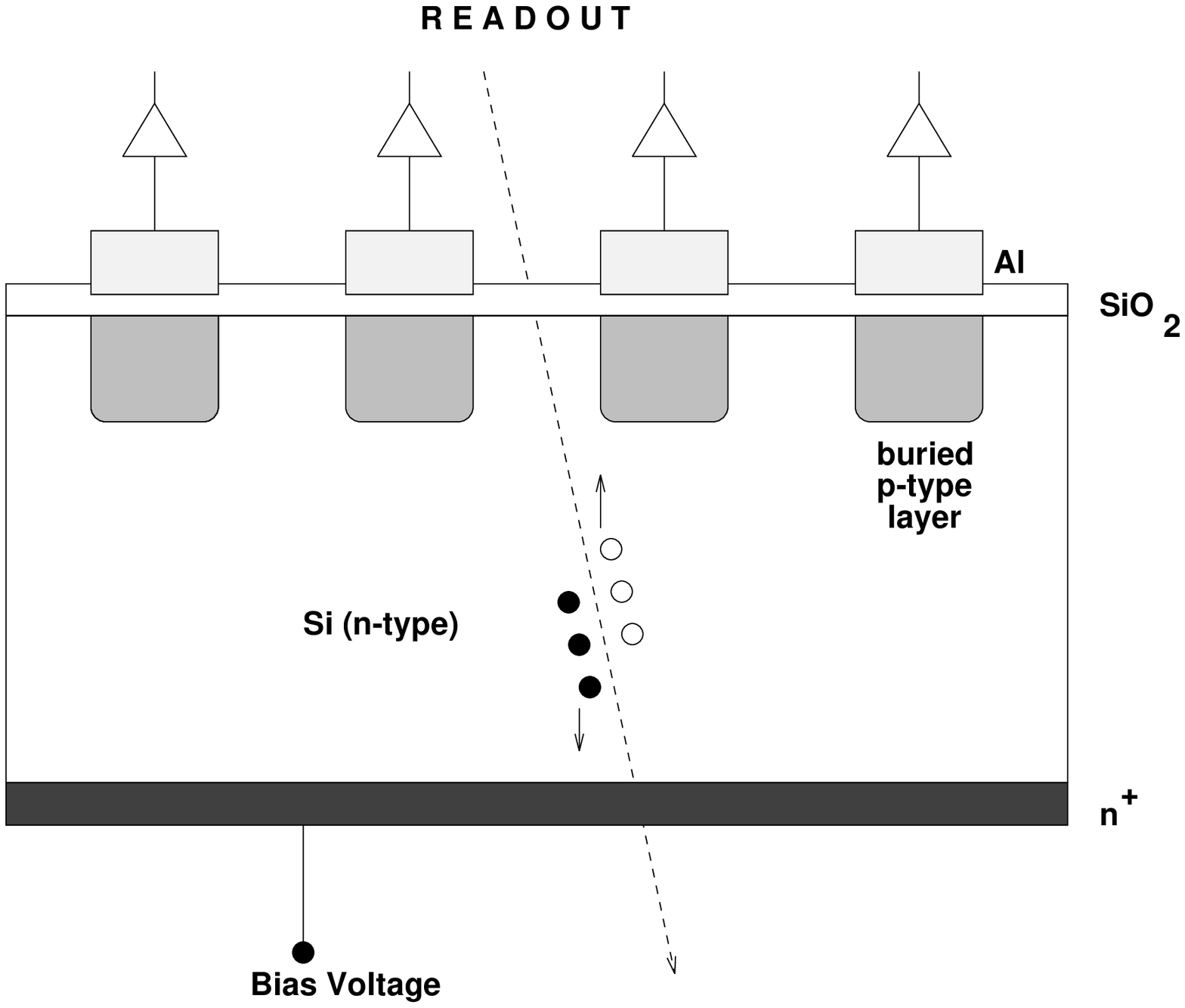}
\caption{Left: Model of a pn-junction in a semiconductor. Right: Schematic
drawing of silicon strip detector.
Both figures taken from~\protect\cite{jurgen:prakash}.}
\label{pnjunction}
\end{center}
\end{figure}

The detector allows to measure with a precision of down to a few $\mu\mbox{m}$
the one-dimensional position of a passing charged track.  Newer devices,
the so-called pixel detectors, measure a two-dimensional position.
The detector uses as basic detection device a pn-junction, shown in 
fig.~\ref{pnjunction} left,
a diode which is operated in blocking direction with a sufficiently high 
voltage so that the entire device is depleted, e.g.\ there are no free
electron-hole pairs (also called charge carriers).  In one device several
(thousands) of these pn-junctions are operated, arranged
in parallel strips.  Should a charged particle pass through the detector
(see fig.~\ref{pnjunction}~right), new electron-hole pair are created
and one of the carrier types will drift toward the nearest strip. In 
Silicon, the energy loss $dE/dx\approx 3.8\,\mbox{MeV/cm}$, and the energy 
needed to create one electron-hole pair is 
$3.6\,\mbox{eV}$\footnote{The band gap in Silicon is only $1.1\,\mbox{eV}$,
but Silicon is an indirect semiconductor.}, so in a
typically $300\,\mu\mbox{m}$ thick detector about $3\cdot10^4$ pairs will be
created. 

The construction of the detector itself seems to be under control today.
There are several companies available which will produce the silicon
detector with a well understood process.  The smallest strip distance
used today is $10\,\mu\mbox{m}$, so that the structure is actually much
simpler than the achieved sub-micron structures in todays computer chips.
The real challenge in these detectors is the readout:  Imagine a 
$5\,\mbox{cm}\times5\,\mbox{cm}$ detector with $10\,\mu\mbox{m}$ strip 
distance: 5000 strips with their small signals have to be read out. Every
single strip needs a preamplifier, and some kind of signal detection like
a discriminator, otherwise noise will overwhelm the data acquisition.
To reduce the number of cables (anyway, how to have a cable 
every $10\,\mu\mbox{m}$?)\ it would be nice to chain several channels 
together, at best even all 5000.  The chips should then be clever enough only
to send a strip number to the data acquisition, e.g.\ the signal gets 
digitized and zero suppressed already at the detector.

A system like this, called SVX~\cite{jurgen:lblsvx}, 
was developed about 10~years ago by LBL for collider experiments (CDF),
and also used 
in WA89~\cite{jurgen:wa89silicon} and 
SELEX~\cite{jurgen:prakash,jurgen:selexsilicon}.  Never version are at use
now in D0~(SVXII) and CDF~(SVXIII).

\begin{figure}
\begin{center}
\includegraphics[width=0.6\textwidth]{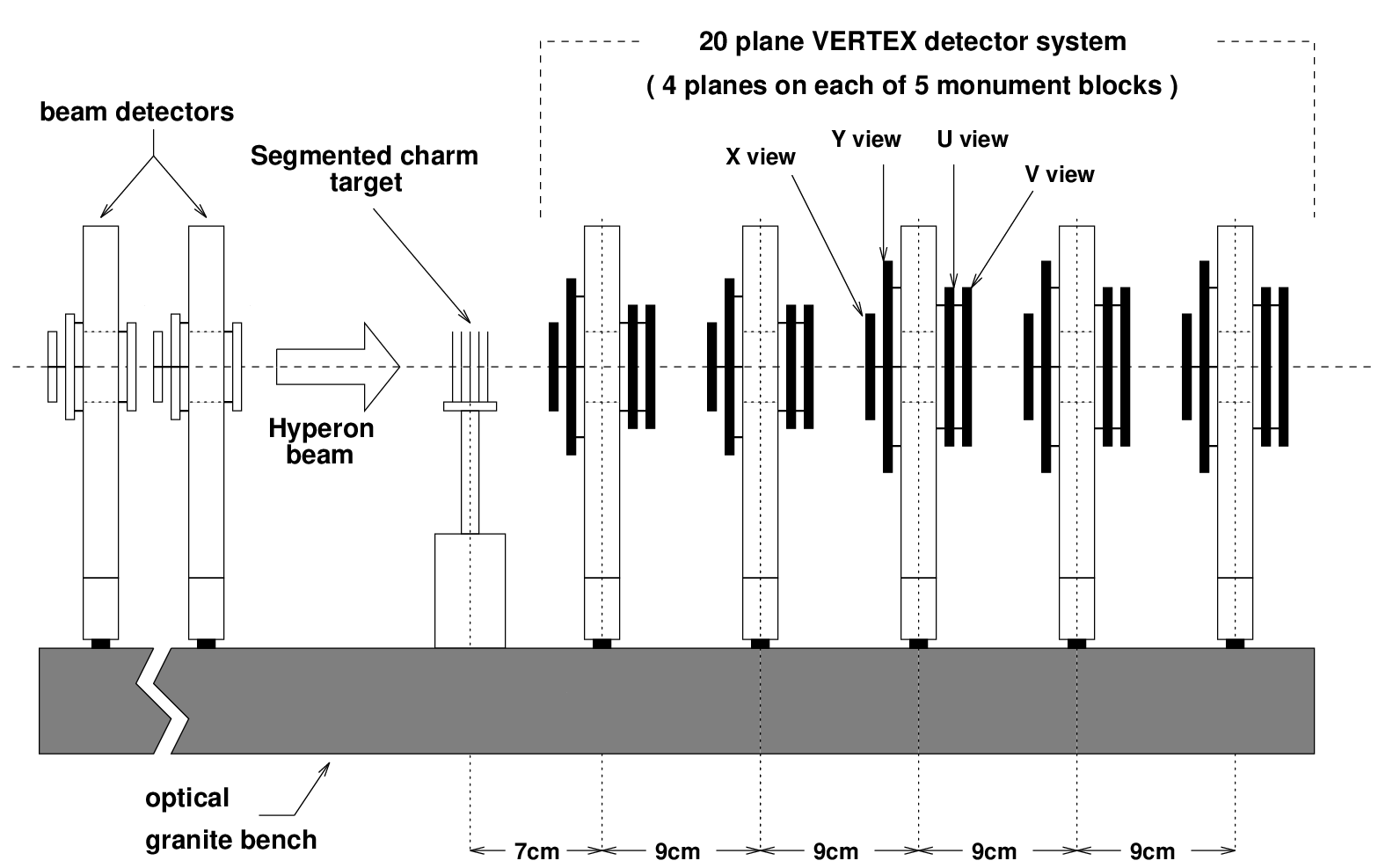}
\hfill
\includegraphics[width=0.39\textwidth]{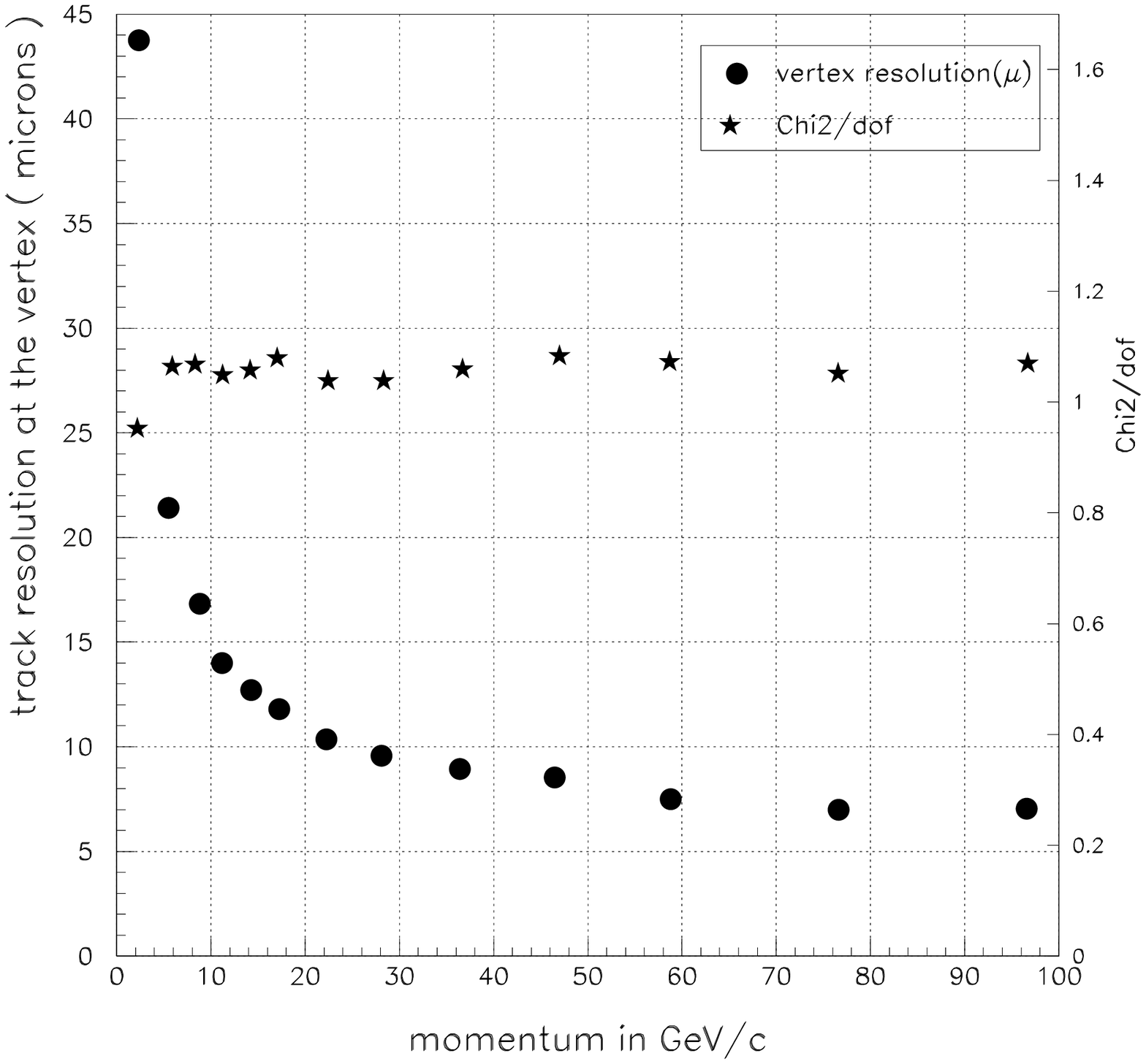}
\caption{
Left: Layout of the SELEX vertex detector. After the target is a total
of 20~planes with $20\,\mu\mbox{m}$ and $25\,\mu\mbox{m}$ 
strip distance in 4 orientations.
Right: Mean $\chi^2/\mbox{dof}$ and vertex track resolution as a function
of momentum.
Figure taken from~\protect\cite{jurgen:prakash}.}
\label{jurgen:siliconlayout}
\label{jurgen:silresol}
\end{center}
\end{figure}

The layout of a typical fixed target vertex detector is shown in 
fig.~\ref{jurgen:siliconlayout}.
Tracks originating from the targets are traversing the silicon planes 
oriented in 4 different orientations (rotated by $45^\circ$) to allow
the reconstruction of tracks in space.  They eventually get fit to form
a vertex, and the obtained resolution is shown in
fig.~\ref{jurgen:silresol} (right).
At high momentum the resolution is limited by the strip distance, but at
lower momentum multiple scattering becomes more and more important.
Nevertheless, the fit takes all error contributions correctly into account,
as seen from a constant $\chi^2=1$ for all momenta.  There is a lesson
to learn:  more detectors is not always good.

\subsection{Other Silicon Detectors}
Other silicon detectors work on the same basic principles.   Double-Sided
strip detectors have a second strip structure on the back side of the
silicon waver, with additional dotations to achieve the diode structure.
The strips are usually, but not always, under $90^\circ$. The readout is more
complicated, two orthogonal sides have to have readout contacts, and in general
double sided silicon is noisier and more difficult to operate.  The signal is
still not a real 3d information, pattern recognition algorithms have to be
used to assign hits in a multi-track environment.  The main advantage is
the reduction of multiple scattering, since one waver can be used to measure
two projections.

Silicon Drift detectors follow the same idea as wire drift chambers: use
the time information to reduce the number of readout channels. This kind
of detector will be used in the ALICE detector~\cite{sildrift}.

Silicon Pixel Detectors have, as says the name, no strips, but usually
rectangular pixels, and give real 3d information of the passing tracks.
The bonding process for the readout electronics is very complicated, but
doable, as the prototypes for the BTeV experiment show~\cite{btev}.
The real 3d information is a key ingredient for the information to
be used in the trigger.

\section{Photomultipliers}
Photomultipliers convert light (photons) via the photoelectric effect
(see section~\ref{photoeffect}) to an electrical signal.  The original
photo-electrons (in the most extreme case only one) is multiplied
by accelerating the electron(s) in an electric field, hitting a metallic
surface and thereby releasing more electrons, in several stages; 
multiplication factors of $10^5$ or more can be obtained.

The efficiency to detect a photon depends on its wavelength and on
the photo-cathode material,
and has a typical maximum of up to 25\%.
\begin{figure}[hbt]
\includegraphics[width=0.44\textwidth,clip]{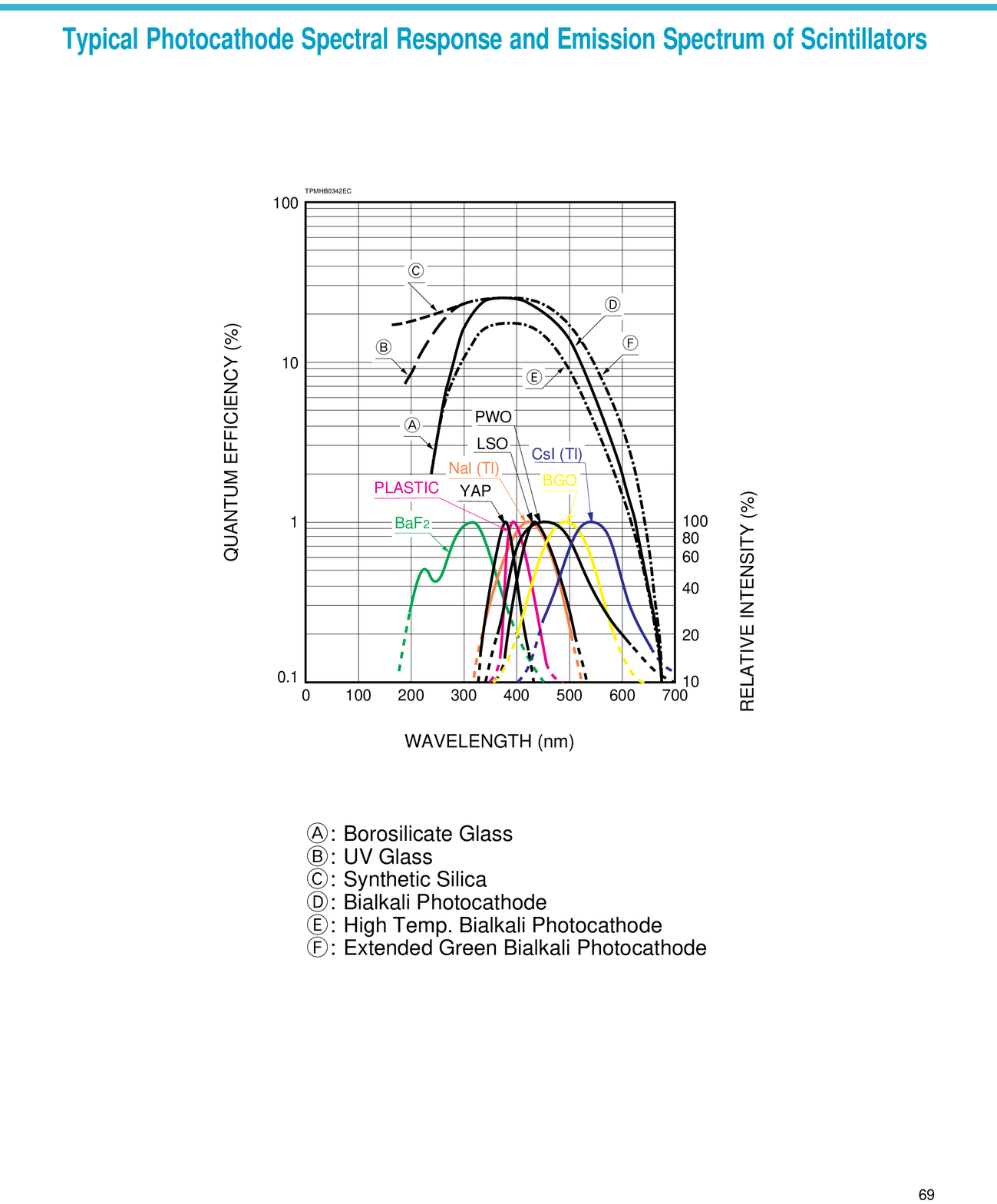}
\hfill
\includegraphics[width=0.54\textwidth]{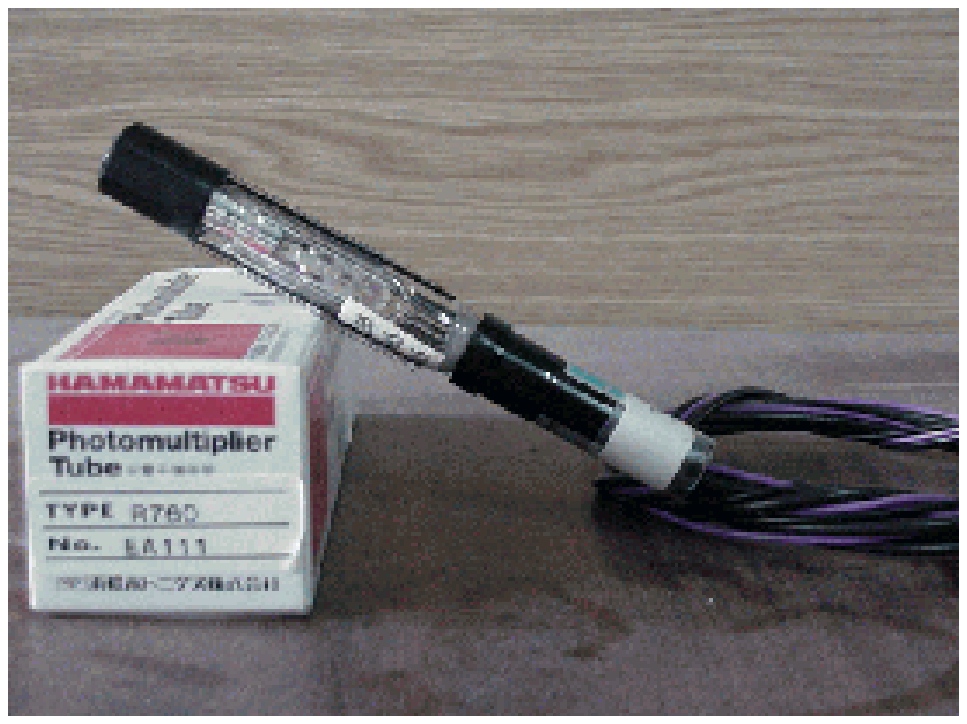}
\caption{Left: Emission spectra for different photon sources, with
the quantum efficiency for different photo-cathode
materials~\cite{hama}.
Right:~Photo of a Hamamatsu R760 1/2\,inch photo-multiplier tube.}
\label{pm-qe}
\end{figure}
The photo-cathode material has to be adopted to the emission spectrum
of the photons to be detected, as shown in fig.~\ref{pm-qe}.

Phototubes are distinguished by their diameter
(available from 3/8\,inch to $\approx 40\,\mbox{cm}$), and the number
of multiplication stages (between 5 and 15). The entrance window
can be on top (``head on'') or on sides (``side on'').
Good summaries about photomultiplier properties can be found
in~\cite{hama,photonics}.

\section{Energy Measurement: Calorimeters}
The shower of an electromagnetic or hadronic cascade has to be totally
absorbed.  If not, the resolution will be reduced due to fluctuations
in the leakage.
For electrons and photons (section~\ref{emshower})
the sizes needed are governed by the
Radiation Length $X_0$ (eq.~\ref{radlen}), 
while for hadrons (section~\ref{hadronshower})
they are governed by the Interaction Length $\lambda_I$.
Different optimizations have to be adopted for electrons/photons and hadrons.
In addition, for a hadron calorimeter we have to take into account
the electromagnetic part of the showers due to $\pi^0$'s, which requires the
concept of ``compensation''~\cite{wigmans}.
A linear dependence on the original particle energy of the output signal
can usually be achieved. The relative energy 
resolution is $\sigma(E)/E\propto 1/\sqrt{E}$, while the shower maximum is 
a logarithmic function of the energy.

\subsection{Electromagnetic Calorimeters}
We can distinguish two basic designs: A 
sampling calorimeter, which we will discuss in more detail in 
section~\ref{hadcal}, and a homogeneous calorimeter.
The latter comes in two groups:  one using the scintillation
properties of matter (section~\ref{scint}), for example CsI crystals,
and one measuring the amount of Cherenkov light (section~\ref{cheren})
produced in a dense but transparent material like lead-glass.
The lateral segmentation is about one Moliere Radius $R_M$ (eq.~\ref{moliere}),
and the depth has to be sufficient to contain the full shower.
Each block of material has a photomultiplier attached at the end; in the
case of the ionization loss measurements, ionization chambers are immersed
into the liquids.
Successful Calorimeters include WA89 Lead-class calorimeter~\cite{wa89lead},
the KTeV CsI calorimeter~\cite{ktev}, and the BGO Calorimeter of L3~\cite{l3}.

\subsection{Sampling Calorimeters}
\label{hadcal}
A sampling calorimeter has 2 parts: 
Something ``heavy'', inactive, like
lead, uranium, iron (steel), and 
something ``light'', active, 
like plastic scintillator (section~\ref{scint}),
wire chamber (section~\ref{mwpc}), liquid (ionization chamber). 
The two materials 
are either put together is an alternating multilayer sandwich or mixed
together, like lead or iron tiles with scintillating fiber readout.

The main design choices is how to use the thicknesses, relative and absolute,
of the two materials.
Problems to consider are:
The inactive part absorbs some of the produced shower particles, and
the absorption fraction is different for hadrons and electrons;
some part of the hadronic response is lost due to neutron absorption;
all hadronic showers have an electromagnetic component due to the
$\pi^0$'s produced in the cascades, and the obtained responses
e.g.\ the physics effects, are different.

The electromagnetic part was understood long time ago 
and good simulation software (EGS) was and is available as part of the
GEANT (see section~\ref{software}) package.
Only in the 1980's, first systematic studies for hadronic responses were
performed.
The first tries led to a 
Uranium / Scintillator sandwich calorimeter, of which a first prototype
had a surprisingly (for
that time) good resolution.  The only partly
correct explanation brought up was
that fission gives back some of ``lost'' neutron energy.  As was found out
later, the first prototype was ``compensated'' by accident.

In the mid-1980's, systematic studies on hadronic shower profiles and
absorptions were started by Wigmans et al.  A good account of this history
is the book by Wigmans~\cite{wigmans}.  The results of these studies was that
{\it any} material combination can be tuned to give identical
responses to electrons and hadrons (``compensate''), due to
the different dependencies on $Z$ of the physics processes involved.
The best resolution for hadrons can only be achieved if the
calorimeter is compensated.
This understanding helped to improve the resolution for hadronic showers, which
was before typically$\sigma_E/E \approx 100\,\%/\sqrt{E}$
to the best achieved values of $30-35\,\%/\sqrt{E}$.

\section{Some words about Electronics}
Electric signals from detectors are small and fast.  This requires
high frequency, high bandwidth, low noise electronics, which are not
totally compatible requirement to easily design the electronics.
In addition, the detector capacitance is usually big, and noise is
often $\propto C^2$. All noise sources 
(shot noise (leakage current), thermal noise, $1/f$ noise)
have to be analyzed with great care.
Stray capacitances and inductances on printed circuits are significant
 at high frequencies ($\geq 100\,\mbox{MHz}$) and have to be taken
into account during the layout of a board, as well as the finite
signal speed ($\sim 5\,\mbox{cm/nsec}$).  All these problems indicated here
require a much more detailed study as can be provided here.  A first start
for further reading is the electronics section of the
Particle Data Book~\cite{pdg}.

\section{Some words about Simulation}
\label{software}
Detectors are complex, complicated and expensive to build.
Within one sub-detector and different sub-detectors of an experiment
interferences, which seem to be mostly destructive, are possible.
Long before building even prototypes, simulations of parts or full
behavior and responses of sub-detectors or full detectors is absolutely
necessary.  The software tool used for these tasks is GEANT~\cite{geant}.
This software package includes models all interactions of all particles, which
can selectively be switch on or off.  The distribution of material is given
as a start, and GEANT is doing everything else.
The original versions were written in FORTRAN, the preferred language
of particle physicists. The documentation is extensive, and also can
serve as a summary of the modeling of physics processes.
Unfortunately, for complex geometries simulating can be very slow, and it
is often necessary to parametrize the response of detectors.

\newpage
\leftline{\Large\bf Part III: Particle Identification}
\vspace{0.3cm}
Here we will discuss as an example for the application of the
detectors described up to now, the identification of particles.
We will discuss in more detail transition radiation detectors and Cherenkov
detectors.

To identify neutral particles, we can only measure their total energy
and, if no charged track is pointing to the signal
in the calorimeter, conclude that this was a neutral particle.  Usually
not too many possibilities are left; for example in a hadron calorimeter,
the only choices are neutrons or $K^0_L$'s.

In an electromagnetic calorimeter, one can measure in addition ``$E/p$'',
e.g.\ the deposited energy over the previously measured momentum of
a charge particle.  If
$E/p=1$ we have a (relativistic) electrons, $E/p<1$ for hadrons, and
if $E/p$ is compatible with minimum ionization, we have a muon.

Long-lived neutral particles like the hyperons $\Lambda^0$ 
and $\Xi^0$, and short lived (but still weakly decaying) particles ($\tau$, 
charm, beauty), we have to measure the 4-vector of all 
decay products to be able calculate the invariant mass of the final state.

The second major applications is
the identification of beam particles for Fixed Target experiments.
In both cases, the momentum of the particle is known, in the case of
neutral particles by guessing, for charged particles by direct measurement
in the beam line  or by a magnetic spectrometer.  With this in mind,
particle identification reduces to measure some additional quantity,
like to total energy with a calorimeter, or some velocity ($\beta$) dependent
effect like 
time-of-flight (section~\ref{tof}), 
$dE/dx$ (eq.~\ref{bethe-bloch}),
Cherenkov (section~\ref{cheren}) or 
Transition Radiation (eq.~\ref{transrad}).

\section{Time--of--flight (TOF)}
\label{tof}
This is the simplest method for identifying particles: Put two
scintillation (or gas) counters at a known distance and
measure the time difference between the two signals.  Time resolutions
of about $150\,\mbox{psec}$ can be achieved, and with the maximum
distance possible between the two detectors
($\approx 10\,\mbox{m}$ for measuring decay products,
$\approx 100\,\mbox{m}$ in a beam line) 
kaons and pions can be separated up to a few $\mbox{GeV}/c$.
At higher rate and/or more than one particles hitting the 
same detector elements the time difference measurement is ambiguous
and the method will not work anymore.

\section{Transition Radiation Detectors (TRD)}

The physics behind these detectors was discussed in section~\ref{tr}.
The detection of Transition Radiation is complicated by the fact that
the X-rays are emitted under a small angle with respect to the particle
track (eq.~\ref{transrad}),
so in a detector the X-ray and the $dE/dx$ signal are seen together and
are of
the same order of magnitude.  In addition, the $dE/dx$ signal is Landau
distributed, forcing the use of several ($10-30$) 
radiator stacks and chamber (with
heavy gases like Xenon or Krypton) units
to be able to distinguish  the two
signals\cite{wa89trd,kolya,errede,stefantrd}.

Some detectors exploit the
fact that the ionization due to the X-ray is more point-like (photoelectric 
effect), as opposed to the more equally distributed $dE/dx$ signal.
The Charge Integration method measures the total charge of the
ionization in every unit, and counts how many chambers saw a total 
charge larger than some given threshold.
The Cluster Counting method employs the different spatial ionization 
distribution of the two source (point for X-ray, line for $dE/dx$)
to separate them.
The obtained signals (either clusters or charge) are compared with
a likelihood method to differentiate particle hypothesis, with known momentum.
The most probable hypothesis is selected as the result of the identification.

The most extreme and sophisticated usage of TRD technology is planned
for the ATLAS experiment at the LHC. They will use more than 370,000
straw tubes as transition radiation detector and in addition
use the obtained information for tracking~\cite{dolgoshein}.

\section{Cherenkov Detectors}

\subsection{Threshold Cherenkov Detectors}
With a suitable medium, which has an refractive index where one of the
particles is above and the other below the threshold
(eq.~\ref{cherenkovthres})
to produce Cherenkov
light, these detectors are used in fixed target beam lines to
separate two particles with the same (fixed) momenta.

For the identification of decay products over a wider momentum range,
and/or if more than two particles have to be distinguished, several
counters at different thresholds have to be employed. Since each of
them is typically several meters long, in practice not more than
three are used. The combination of signals for different momentum
ranges yields a decent identification probabilities.

\subsection{Ring Imaging Cherenkov Detectors (RICH)}
\label{rich}
Even though the basic idea of determining the velocity of charged particles
via measuring the Cherenkov angle was proposed in 1960~\cite{roberts},
and in 1977 a first prototype was successfully operated~\cite{seguinot},
it was only during the last decade that Ring Imaging Cherenkov (RICH) 
Detectors 
were successfully used in experiments.  A very useful collection of 
review articles and detailed descriptions can be found in the proceedings
of four international workshops on this type of
detectors, which were held in 
1993 (Bari, Italy)~\cite{richworka}, 
1995 (Uppsala, Sweden)~\cite{richworkb}, 
1998 (Ein Gedi, Israel)~\cite{richworkc}, and 
2002 (Pylos, Greece)~\cite{richworkd}, respectively.
In 2004 the workshop will be held in Mexico~\cite{rich2004}.

By measuring the Cherenkov angle $\theta$ one can in principle
determine the velocity
of the particle, which will, together with the momentum $p$ obtained via
a magnetic spectrometer, lead to the determination of the mass and therefore
to the identification of the 
particle.

Neglecting multiple scattering and energy loss in the medium, all the Cherenkov
light (in one plane) is parallel, and can therefore be focused (for small
$\theta$) with a spherical mirror (radius $R$) onto a point.
Since the emission is symmetrical in the azimuthal angle around the
particle trajectory, this leads to a ring of radius $r$ in the
focus, which is itself a sphere with radius $R/2$. The  radius $r$ is given by 
\begin{equation}
r = {{R}\over{2}} \tan\theta
 \approx {{R}\over{2}}\sqrt{2 - {{2}\over{n}}\sqrt{1+{{m^2 c^2}\over{p^2}}}}
\end{equation}

\subsubsection{A Short History of RICHes}
After the first successful laboratory tests in the 1970's~\cite{seguinot},
at the beginning of the 1980's the first generation of Ring Imaging
Cherenkov Detectors were built, with mixed results. Examples are
the CERN Omega RICH, used by WA69 and WA82~\cite{omegarich}, the UA2 RICH
at CERN, and 
the E653 RICH at Fermilab.

A second generation RICH detectors were developed and employed at the
end of the 80's and beginning of the 90's, with the
positive and negative experiences
from the first generation incorporated.
Examples are an upgraded Omega RICH, used by WA89 and WA94~\cite{omegarich},
Delphi, CERES (all at CERN), and the SLD--GRID at SLAC.
All these detectors had significant startup problems to overcome,
but eventually
they all worked very well and contributed to the physics analysis of the
experiments.

The third generation, build and used mid-end 90's, finally worked without too
many problems from the beginning and all of them played an important role
in the physics analysis. Examples are the SELEX RICH at Fermilab, 
and Hermes and Hera-B at DESY.

Since the mid-90's, the RICHes are an established detector type, and are
currently employed in recent experiments:
BaBar--DIRC (SLAC), PHENIX and STAR (Brookhaven-RHIC), CLEO--III (Cornell), 
COMPASS (CERN).

There is a long list of future experiments, who plan to or are already
building RICHes: ALICE, LHCB (CERN), BTeV, CKM (Fermilab), and many more.
Also in other fields like nuclear physics (GSI) and in astroparticle
physics experiments technology developed for RICHes is used.

\subsubsection{General Details for RICHes}

Since the number of photons is $\propto\lambda^{-2}$ (eq.~\ref{jurgen:dndl}),
most of the light is
emitted in the VUV range.  To fulfill equation~\ref{jurgen:cherenkovangle},
the refractive index has to be $n>1$, so there will be no Cherenkov radiation
in the x-ray region.  Also it is very important to remember that $n$ is a 
function of the wavelength ($n=n(\lambda)$, chromatic dispersion)
and most materials have an
absorption line in the VUV region, where $n\to\infty$.
Since usually the
wavelength of the emitted photon is not measured, this leads to a smearing
of the measured ring radius, and one has to match carefully the wavelength
ranges which one wishes to use:  Lower wavelengths gives more photons, but
larger chromatic dispersion.

A very useful formula is obtained by integrating eq.~\ref{jurgen:dndl}
over $\lambda$ (or $E$),
taking into account all efficiencies etc., obtaining a formula for
the number of detected photons $N_{\rm ph}$~\cite{seguinot}:
\begin{equation}
N_{\rm ph} = N_0 L \sin^2\theta
\end{equation}
where $N_0$ is an overall performance measure (quality factor) of the detector,
containing all the details (sensitive wavelength range, efficiencies), and
$L$ is the path length of the particle within the radiator. A `` very good''
RICH detector has $N_0=100\,\mbox{cm}^{-1}$, which gives typically around
$10$ to $15$ detected photons ($N_{\rm ph}$) per $\beta=1$ ring.

The usual construction of a RICH detector is to use a radiator length of
$L= R/2$, e.g.\ the path length is 
equal to the focal length; but any other configurations, like
folding the light path with additional (flat) mirrors, are possible.

All the presented arguments
only work for small $\theta$, which is always fulfilled in gases,
since $n$ only differs little from $1$.  Also important is the fact, that,
should the particles not pass through 
the common center of curvature of mirror(s)
and focal spheres, the ring gets deformed to an ellipse or, in more extreme
cases, to a hyperbola. If the photon detector is able to resolve this, and the
resolution is needed for the measurement, these deviations from a perfect
circle have to be taken into account in determining the velocity $\beta$.
In general this effect can be  neglected, and all parallel particles (with
the same $\beta$) will give the same ring in the focal surface, due to the
fact that all emitted Cherenkov light is parallel.  The position of the 
ring center is determined by the angle of the tracks, not by their positions.

In the following, we will describe three RICH detectors used in experiments,
and a new application for RICH detectors for a new experiment.

\subsubsection{The CERN Omega-RICH}

In the middle of the 1980's, first attempts were made to apply the
prototype results obtained by S\'eguinot and Ypsilantis~\cite{seguinot}
to experiments in a larger scale.  One of these attempts was performed
at the CERN Omega facility in the West Hall. Experiments WA69 and WA82
tried to use this detector for their analysis, but only succeeded partly.
An overview about this history can be found in~\cite{omegarich}.
When in 1987 a new experiment, later named WA89, 
was proposed~\cite{jurgen:wa89loi,jurgen:wa89proposal}, an
important part was a necessary upgrade of this detector for the use by
this new experiment.  Two main parts where changed:  New photon detectors
using TMAE as photon sensitive component, and new mirrors to perform the
focusing.  Details about the detector 
can be found 
in~\cite{omegarich,jefthesis,wa89rich}.

The overall layout of the detector shows that 
RICH detectors are basically simple devices:  a big box, some mirrors at the
end, and photon-sensitive detectors at the entrance.
The real challenge is to combine all the parameters together to obtain 
a perfect match for the overall system.

The size of the radiator box 
and the photon detector is given by the angular distribution of tracks which
have to be identified at the location of the detector.  Since usually this
detector is placed behind a magnetic spectrometer, and the momentum spectrum
of the interesting tracks depends on the physics goals of the experiment,
the surfaces to cover have to be determined for every setup and experiment,
usually with Monte Carlo simulations during the design phase of the
experiment.
In the case of WA89~\cite{jefthesis}, the mirror surface needed was
about $1\,\mbox{m}\times 1.5\,\mbox{m}$, much smaller than the 
$4\,\mbox{m} \times 6\,\mbox{m}$ covered by the original Omega-RICH. It was
therefore decided to replace only the central mirrors with smaller,
higher surface quality mirrors to obtain better resolution.

The detector surface was calculated to be $1.6\,\mbox{m}\times0.8\,\mbox{m}$,
with a spatial resolution of a few millimeters for every detected photon.
The pixel size could therefore not be much bigger than also a few millimeters,
leading to about $100000$ pixels in the detector plane.  The solution
was to build drift chamber (see section~\ref{driftchamber}) (TPC) modules, 
covering an area of
$35\,\mbox{cm}\times 80\,\mbox{cm}$, and approximating the focal sphere
with a polygon of 5~modules.  After passing a $3\,\mbox{mm}$ thick 
(to minimize absorption) quartz window, the photons are absorbed by
Tetrakis(dimethylamino)-ethylen (TMAE) molecules, 
converting via photo-effect~(section~\ref{photoeffect})
into a single electron. TMAE is present
with a concentration of about $0.1\,\mbox{\%}$ within the driftgas, which
is otherwise pure Ethan. Due to the use of a quartz window together with TMAE
as photon-sensitive gas the detector is only sensitive in a small
wavelength range between $165\,\mbox{nm}< \lambda <230\,\mbox{nm}$.
TMAE has a very low vapor pressure, so that at ambient temperatures the
molecules are saturated within a gas.  To obtain a short enough conversion
length of around $1\,\mbox{cm}$ (otherwise the conversion would occur
to far away from the focal plane and lead to an additional contribution to
the resolution (parallax)),
the drift gas (Ethan) is led through a bubbler, containing
TMAE liquid at $30^\circ\,\mbox{C}$. 
This means that everything after the bubbler,
e.g.\ the whole detector including radiator box, had to be heated to
$40^\circ\,\mbox{C}$ to avoid condensation.  Other unpleasant properties
of TMAE include a high reactivity with Oxygen, producing highly 
electro-negative
oxides, which will attach an electron easily, changing the drift velocity
by a factor of several thousand, leading to a loss of electrons.  Since
the signal is a single electron (photo-effect!)\ this is catastrophic.  The
counting gas had an Oxygen contents of $<1\,\mbox{ppm}$.

Mostly due to the presence of TMAE,  the operation of this detector was 
not trivial.  All parameters were monitored electronically, and
hardware limits on some critical
parameters (like temperature, Oxygen content of Ethan) lead to a automatic
shutdown of the detector, waiting for an expert to arrive in the experimental
hall.
  
Once the electron was released, it was drifting under the influence of an 
electric field of $1\,\mbox{kV/cm}$ 
(drift velocity $5.4\,\mbox{cm/}\mu\mbox{m}$) upwards or downwards 
over maximal $40\,\mbox{cm}$ toward
$6\,\mbox{cm}$ long counting wires (gold-coated tungsten, $15\,\mu\mbox{m}$
diameter), spaced by $2.54\,\mbox{mm}$.
The two-dimensional 
spatial information about the
conversion point of the photon is obtained with the position of the wire and
the drift time of the electron. In total, 1280 wires were used in the
detector.  An additional complication was that the charged particles itself
were passing through the chambers, leaving a $dE/dx$ signal of several
hundred electrons, which is to be compared to the single electron which is
our signal.  This leads to increased requirements for the wire chambers
(sensitive, e.g.\ sufficient multiplication, to single electrons,
but no sparking with several hundred electrons) and to the preamplifier
electronic (not too much dead time after a big pulse).

The overall resolution allowed the separation of pions and kaons up to 
a momentum of about $100\,\mbox{GeV/}c$, which was exactly the design goal.
This led to a good number of physics 
results (about 15 papers),
which would not have been possible to obtain without the RICH detector.

\subsubsection{The SELEX Phototube RICH Detector}

At Fermilab a new hyperon beam experiment, called SELEX, was proposed
in 1987~\cite{propselex}.  The key elements to perform a successful
charmed-baryon experiments are 1)~a high resolution silicon vertex detector
and 2)~a extremely good particle identification system based on RICH.
During the following years, a prototype for the SELEX RICH was constructed and
tested successfully~\cite{selexproto}, 
based in some part on experience gained
by our Russian collaborators~\cite{sphinx}.
The real detector was constructed in 1993-1996, ready for the SELEX 
data taking period from July~1996 to September~1997.
Details and performance descriptions of this detector can be
found in~\cite{selexrich}.

\begin{figure}[htb]
\begin{center}
\includegraphics*[width=15cm]{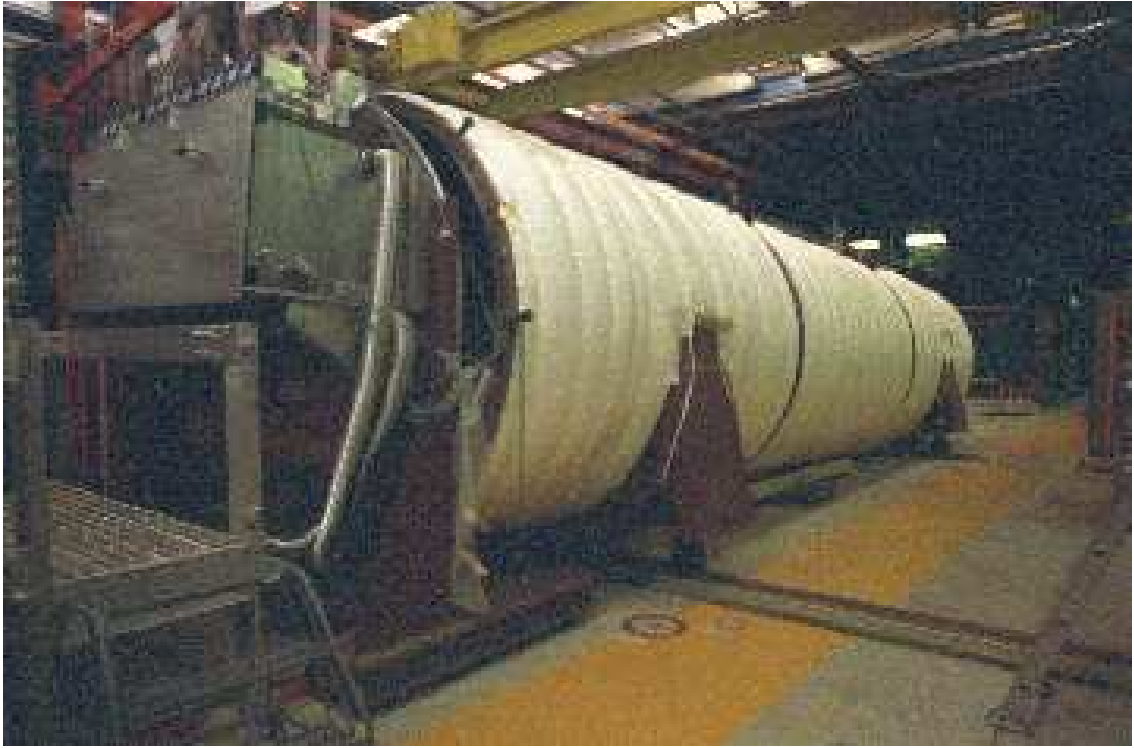}
\end{center}
\caption{The SELEX RICH detector.  The vessel has an overall
length of a little over $L=10\,\mbox{m}$, the mirrors have a radius of
$R=1982\,\mbox{cm}$, and the light tight box containing the photomultipliers
is clearly seen on the left.
 The whole vessel is tilted by $2.4^\circ$ to avoid that
the particle trajectories go through the
photomultipliers.}
\label{jurgen:selexvessel}
\end{figure}
A photo of the detector is shown in fig.~\ref{jurgen:selexvessel}.
The radiator gas is Neon at atmospheric pressure and room
temperature, filled into the vessel with a
nice gas-system~\cite{selexgas}: 
First the vessel is flushed for about 1~day with 
${\rm CO}_2$ (a cheap gas). After this 
the gas (mostly ${\rm CO}_2$ and little air) is pumped in a closed system
over a cold trap running at liquid Nitrogen temperature, freezing out
${\rm CO}_2$ and the remaining water vapor.  At the same time Neon gets filled
into the vessel to keep the pressure constant.  This part of the procedure
takes about 1/2~day, and the vessel contains afterward only Neon and 
about $100\,\mbox{ppm}$ of Oxygen which is removed by pumping the gas
over a filter of activated charcoal for a few hours, ending with an Oxygen
contents of $<10\,\mbox{ppm}$ in the radiator.  After this all valves were
closed and the vessels sits there for the whole data taking of 
more than 1~year at a slight ($\approx 1\,\mbox{psi}$)
over-pressure.

\looseness=-1
The mirror array at the end of the vessel is made of $11\,\mbox{mm}$
low expansion glass, polished to an average radius of 
$R=(1982\pm 5)\,\mbox{cm}$, coated with Aluminum and a thin over-coating
of ${\rm MgF}_2$, which gives $>85\,\mbox{\%}$ reflectivity at 
$155\,\mbox{nm}$.
The quality of the mirrors was measured with the Ronchi
technique~\cite{ronchi} to assure a sufficient surface quality of the
mirrors. The total mirror array covers $2\,\mbox{m} \times 1\,\mbox{m}$
and consists of 16~hexagonally shaped segments.  The mirrors are fixed with
a 3-point mount consisting of a double-differential screw and a ball bearing
to a low mass honeycomb panel.  The mirrors are mounted on one sphere, and
were aligned by sweeping a laser beam coming from the center of curvature
over the mirrors.

The photo detector is a hexagonally closed packed $89\times 32$ array of
2848 half-inch photomultipliers.
In a $3\,\mbox{in.}$\ thick aluminum plate holes are drilled from both sides,
a $2\,\mbox{in.}$ deep straight hole holds the photomultiplier, and 
a conical hole on the radiator side holds aluminized Mylar Winston cones, 
which form on the radiator side hexagons, leading to a total coverage of the
surface.  The 2848~holes are individually sealed with small quartz windows.
For the central region of the array, a mixture of
Hamamatsu R760 (fig.~\ref{pm-qe})
 and
FEU60 tubes were used, in the outside rows only FEU60 tubes are present.
The nearly 9000~cables (signal, hv, ground) are routed to the bottom (hv)
or top, where the signal cables are connected to 
preamp-discriminator-ecl-driver hybrid chips and finally readout via
standard latch modules\footnote{Since the phototubes are detecting single
photons, no ADCs are necessary.}.

The SELEX RICH demonstrated a
 clear multitrack capability and low noise.
To analyze an event, the ring center is predicted via the known
track parameters, and a likelihood 
analysis~\cite{partid} for different hypothesis
(the momentum is known!)\ is performed to identify the particle. 
The final performance of this detector was 
evaluated~\cite{selexrich}.
The detector is nearly $100\,\mbox{\%}$ efficient for protons;
even below the proton threshold the efficiency is above $90\,\mbox{\%}$. 
In the SELEX offline analysis, the RICH is one of the first cuts applied
to extract physics results.

Up to date, SELEX published 10~papers with physics results from charm
and hyperon physics, three of them in PRL.

\subsubsection{DIRC at BaBar}
BaBar is an experiment at SLAC, one of the so-called B-Factories, which 
are measuring in great detail ${\cal CP}$-violation parameters in B-Mesons.
They employ a RICH detector which they call DIRC
(Detection of Internally Reflected Cherenkov light)~\cite{babar}.

The radiator in this case are quartz bars, which, by internal reflection,
transport the Cherenkov light to one side (the other side is mirrored).
A standoff box, filled with water, expands the light and couples it to
about 10,000 photomultiplier tubes.
Due to the multiple internal reflections,
the primary images are not rings, but present a more complicated pattern.
Nevertheless, by considering in addition the arrival time of the photons
at the PMTs, the Cherenkov angle can be reconstructed.

The performance of the DIRC is also very good. They reported
high kaon efficiency and a low miss-identification probability.
More information can be found in~\cite{schwiening}.

\subsubsection{Future of RICHes with a Mexican touch: CKM}

In 2001, a new experiment called CKM~\cite{ckmprop} received
Stage~I approval at Fermilab.
The goal of the experiment is to  measure the branching ratio for
$K^+ \to \pi^+ \nu\bar\nu$ to an accuracy of $10\,\mbox{\%}$ (SM prediction
is $10^{-10}$) to measure the CKM matrix element $V_{td}$,
contributing to test the Standard Model
hypothesis that a single phase in the CKM matrix is the sole
source of ${\cal CP}$ violation.
To withstand the high expected physics background, the
experiment will use, in addition to a conventional magnetic spectrometer,
a velocity spectrometer consisting of two phototube RICH detectors, one
to measure the incoming $K^+$, the second the outgoing
$\pi^+$~\cite{ckmrich}.  The experiment is expected to take data
in the second half of this decade.

The expected ring radii for $K^+$ and $\pi^+$ from the beam, measured
in the Kaon RICH, 
are very well separated, so the usual
resolution question is not appropriate to ask.  The real question is:
how Gaussian is the response function?  The result, checked with SELEX
single track data, 
using the standard SELEX algorithm, is that the response function is
Gaussian over nearly 5~orders of magnitude.

\subsubsection{Testing Cherenkov Mirrors}

The HEP group in San Luis
Potos\'{\i}, Mexico is involved in the design, construction, and testing of
parts of for the CKM RICH detectors.
We will mention here especially the construction
and testing of mirrors. We apply the Ronchi method~\cite{ronchi}.
The radius of a mirror is calculated from the distance of lines from
a grating projected to the mirror surface.
We are currently testing the first prototype mirrors
produced in Mexico.%\footnote{CIO, Le\'on, Gto.}.  

\section{Summary Particle Identification}
\begin{figure}
\begin{center}
\leavevmode
\epsfxsize=\hsize
\epsffile{lecturenotes-grupen.ps_page_45}
\end{center}
\caption{  }
\label{sumpartid}
\end{figure}
As a summary, we show in fig.~\ref{sumpartid} how to use different
detectors to separate particles.  Incidentally, in a fixed target 
experiment in a linear sense, and in a collider detector radial,
this is the order of detectors the particles pass through.

\section{Acknowledgment}
The author thanks the organizers of this
school for the invitation and the privilege to give a course on his
favored topic.


\begin{thebibliography}{99}

\bibitem{grupen}
Claus Grupen: {\sl Particle Detectors}. Cambridge University Press (2000).

\bibitem{klein}
Konrad Kleinknecht: {\sl Detectors for particle radiation}.
Cambridge University Press, 2nd edition (1998).

\bibitem{fernow}
Richard C. Fernow: {\sl Introduction to experimental Particle Physics}.
Cambridge University Press (1986).

\bibitem{wigmans} 
Richard Wigmans: {\sl Calorimetry}. Oxford Science Publishing (2000).

\bibitem{icfa}
C.W.~Fabjan, J.E.~Pilcher (Eds.):
{\sl Instrumentation in Elementary Particle Physics. Proceedings, ICFA School,
Trieste, Italy, June 8-19, 1987}.
World Scientific (1988).\\
J.C.~Anjos, D.~Hartill, F.~Sauli, M.~Sheaff (Eds.):
{\sl Instrumentation in Elementary Particle Physics. Proceedings, 3rd ICFA 
School, Rio de Janeiro, Brazil, July 16-28, 1990}.
World Scientific (1992). \\
G.~Herrera Corral, M.~Sosa Aquino (Eds.):
{\sl Instrumentation in Elementary Particle Physics. Proceedings, 7th ICFA
 School, Leon, Mexico, July 7-19, 1997}.
AIP Conference Proceedings 422 (1998).\\
S.~Kartal, (Ed.):
{\sl Instrumentation In Elementary Particle Physics. Proceedings, 8th ICFA
 School, Istanbul, Turkey, June 28-July 10, 1999}.
AIP Conference Proceedings (2000).\\
L.~Villase\~nor, V.~Villanueva, (Eds.):
{\sl Instrumentation in Elementary Particle Physics. Proceedings,
1st ICFA Instrumentation School, Morelia, Mexico, November 18-29, 2002}.
AIP Conference Proceedings 674 (2003). 

\bibitem{pdg}
K.~Hagiwara et al.: Physical Review {\bf D66} (2002), 01001-1.
{\tt http://pdg.lbl.gov}

\bibitem{pdg04}
Particle Data Group, to be published (2004).
{\tt http://pdg.lbl.gov}

\bibitem{peptpc}
D.R.~Nygren, J.N.~Marx: {\sl The Time Projection Chamber}.
Phys.\ Today {\bf 31}, 46 (1978).

\bibitem{jade}
J.~Heintze:
{\sl The Jet Chamber of the JADE Experiment}.
Nucl.\ Instrum.\ Meth.\ {\bf 196} 293-297 (1982).

\bibitem{opal}
H.M.~Fischer et al.: {\sl The OPAL Jet Chamber Full Scale Prototype}.
Nucl.\ Instr.\ Meth.\ {\bf A252} (1986) 331-342.\\
H.~Breuker et al.:
{\sl Particle Identification with the OPAL Jet Chamber in the Region of the 
Relativistic Rise}.
{Nucl.\ Instr.\ Meth.\ }{\bf A260}\,(1987)\,329.\\
R.D.~Heuer, A.~Wagner:
{\sl The OPAL Jet Chamber}.
{Nucl.\ Instr.\ Meth.\ }{\bf A265}\,(1988)\,11-19. 

\bibitem{menchaca}
K.~Michaelian, A.~Menchaca-Rocha, E.~Belmont-Moreno:
Nuclear Instruments and Methods {\bf A356}, 297-303 (1995).

\bibitem{cherenkov}
P.A.~Cherenkov:
{\sl Visible radiation produced by electrons moving in a medium
with velocities exceeding that of light}.
{ Phys.\ Rev.\ }{\bf 52}\,(1937)\,378.\\
Pavel A.\ Cherenkov, Il´ja M.\ Frank, Igor Y.\ Tamm, Nobel Price 1958.

\bibitem{francktamm}
I.~Frank, I.~Tamm:
{\sl Coherent visible radiation of fast electrons passing through matter}.
{C.\ R.\ Acad.\ Sci.\ URSS} {\bf 14} (1937) 109.

\bibitem{belle}
A.~Satpathy et al.: BELLE Collaboration, hep-ex/9903045.

\bibitem{selexrich}
J.~Engelfried et al.: Nucl.\ and Instr.\ and Methods {\bf A502} 62-65 (2003).\\
J.~Engelfried et al.:
{\sl The SELEX Phototube RICH Detector}.
{ Nucl.\ Instr.\ and Meth.\ }{\bf A431}\,(1999)\,53-69.
hep-ex/9811001.\\
J.~Engelfried et al.:
{\sl The E781 (SELEX) RICH Detector}.
{ Nucl.\ Instr.\ and Meth.\ }{\bf A409}\,(1998)\,439.\\
J.~Engelfried et al.:
{\sl The RICH Detector of the SELEX Experiment}.
{ Nucl.\ Instr.\ and Meth.\ }{\bf A433}\,(1999)\,149.\\
J.~Engelfried et al.:
{\sl SELEX RICH Performance and Physics Results}.
Nucl.\ Instr.\ and Meth.\ {\bf A502} (2003) 285-288.
Preprint UASLP-IF-02-007, hep-ex/0208046

\bibitem{heintze}
A.H.~Walenta, J.~Heintze, B.~Sch\"urlein:
{Nucl.\ Instr.\ Meth.\ }{\bf 92}\,(1971)\,373.

\bibitem{jurgen:prakash}
P.~Matthew: {\sl Construction and Evaluation of a high Resolution Silicon
Microstrip Tracking Detector and Utilization to determine Interaction
Vertices}. {Ph.D.\ Thesis, Carnegie-Mellon University, Pittsburgh} (1997).  

\bibitem{jurgen:lblsvx}
S.A.~Kleinfelder et al.: Lawrence Berkeley Laboratory Note.

\bibitem{jurgen:wa89silicon}
W.~Br\"uckner et al.:
{\sl Silicon $\mu$-strip detectors with SVX-chip readout}.
Nucl.\ Instr.\ Meth.\ {\bf A348} (1994) 444-448.

\bibitem{jurgen:selexsilicon}
J.~Russ, et al.: {IEEE Trans.\ Nucl.\ Sci.\ }{\bf NS36}\,(1989)\,471.

\bibitem{sildrift}
V.~Bonvicini, et al.: {\sl Silicon Drift Detectors in the ALICE Experiment}.
In: Ayala, Contreras, Herrera (Eds.): {\sl Particles and Fields,
Seventh Mexican Workshop}. AIP Conference Proceedings 531 (2000).

\bibitem{btev}
S.~Kwan et al.:
{\sl Study of Indium and Solder Bumps for the BTeV Pixel Detector}.
FERMILAB-CONF-03-363-E (2003).

\bibitem{hama}
Hamamatsu Photomultiplier Catalog. {\tt www.hamamatsu.com}

\bibitem{photonics}
Photomultiplier Tubes, Principles and Applications.
Photonis ({\tt www.photonis.com}) (2002).

\bibitem{wa89lead}
W.~Br\"uckner et al.:
{\sl The Electromagnetic Calorimeter in the Hyperon Beam Experiment at CERN}.
Nucl.\ Instrum.\ Meth. {\bf A313} 345-356 (1992).


\bibitem{ktev}
R.S.~Kessler et al.:
{\sl Beam test of a prototype CsI calorimeter},
Nucl.\ Inst.\ Meth.\ {\bf A368} 653-665 (1996).

\bibitem{l3}
J.A.~Bakken et al.''
{\sl Results on the Calibration of the L3 BGO Calorimeter with Cosmic Rays}.
Nucl.\ Instrum.\ Meth.\ {\bf A343} 456-462 (1994).

\bibitem{geant}
GEANT -- Detector Description and Simulation Tool, CERN.

\bibitem{wa89trd}
W.~Br\"uckner et al.: {\sl The Transition Radiation Detector in the
Hyperon Beam Experiment WA89 at CERN}.
{ Nucl.\ Instr.\ and Meth.\ }{\bf A378}\,(1996)\,451.

\bibitem{kolya}
N.~Terentyev et al.: 
{\sl E781 Beam Transition Radiation Detector}.
SELEX internal Note H-746.

\bibitem{errede}
D.~Errede et al.: 
{\sl Use of a Transition Radiation Detector in a Beam of High-Energy Hadrons}.
{ Nucl.\ Instr.\ and Meth.\ }{\bf A309}\,(1991)\,386.\\
D.~Errede et al.:
{\sl Design and Performance Characteristics of the E769 Beamline
Transition Radiation Detector}.
{ IEEE Trans.\ Nucl.\ Sci.\ }{\bf 36}\,(1989)\,106.

\bibitem{stefantrd}
S.~Paul: {\sl Particle Identification using Transition Radiation Detectors}.
{CERN-PPE-91-199}.

\bibitem{dolgoshein}
B.~Dolgoshein:
{\sl Complementary particle ID: transition radiation and $dE/dx$ relativistic
rise}.
Nucl.\ Instr.\ and Meth.\ {\bf A433}\,(1999)\,533.

\bibitem{roberts} A.~Roberts:
{\sl A new type of Cherenkov detector for the accurate
measurement of particle velocity and direction}.
{ Nucl.\ Instr.\ and Meth.\ }{\bf 9}\,(1960)\,55.

\bibitem{seguinot} J.~S\'eguinot and T.~Ypsilantis:
{\sl Photo-ionization and Cherenkov Ring Imaging}.
{ Nucl.\ Instr.\ and Meth.\ }{\bf 142}\,(1977)\,377.

\bibitem{richworka}
E.~Nappi, T.~Ypsilantis (Eds.):
{\sl Proceedings of the First Workshop on Ring Imaging Cherenkov Detectors}.
{ Nucl.\ Instr.\ and Meth.\ }{\bf A343}\,(1994).

\bibitem{richworkb}
T.~Ekel\"of (Ed.):
{\sl Proceedings of the Second International Workshop on
Ring Imaging Cherenkov Detectors}.
{Nucl.\ Instr.\ and Meth.\ }{\bf A371} (1996).

\bibitem{richworkc}
A.~Breskin, R.~Chechik, T.~Ypsilantis (Eds.):
{\sl Proceedings of the Third International Workshop on
Ring Imaging Cherenkov Detectors}.
{Nucl.\ Instr.\ and Meth.\ }{\bf A433} (1999).

\bibitem{richworkd}
{\sl Proceedings of the IV.~International Workshop on
Ring Imaging Cherenkov Detectors}.
Nucl.\ Instr.\ and Meth.\ {\bf A502}  (2003).

\bibitem{rich2004}
{\tt http://www.ifisica.uaslp.mx/rich2004}

\bibitem{dispersion} A.~Bideau-M\'ehu et al.:
{\sl Measurement of refractive indices of neon, argon, krypton, and
xenon in the $253.7-140.4\mbox{nm}$ wavelength range}.
{ J.\ Quant.\ Spectrosc.\ Radiat.\ Transfer} {\bf 25} (1981) 395.

\bibitem{omegarich}
H.-W.~Siebert et al.: {\sl The Omega-RICH}.
{ Nucl.\ Instr.\ and Meth.\ }{\bf A343} (1994), 60.

\bibitem{jurgen:wa89loi}
J.~Engelfried et al.:
{\sl A high-statistics experiment on the U(3100) and on 
charmed-strange baryons}.
{Letter of Intent CERN/SPSC/87-8, SPSC/I165} (1987).

\bibitem{jurgen:wa89proposal}
A. Forino et al.:
{\sl Proposal for a new hyperon beam experiment at the CERN SPS using the 
Omega facility}.
{CERN/SPSC/87-43, SPSC/P233} (1987).

\bibitem{jefthesis}
J.~Engelfried: { Ph.D.\ Thesis, Heidelberg University} (1992), unpublished.

\bibitem{wa89rich}
W.~Beusch et al.: {\sl The RICH counter in the CERN hyperon beam experiment}.
{ Nucl.\ Instr.\ and Meth.\ }{\bf A323}\,(1992)\,373.\\
U.~M\"uller et al.:
{\sl The recent performance of the Omega RICH detector in 
experiment WA89 at CERN}.
{ Nucl.\ Instr.\ and Meth.\ }{\bf A371}\,(1996)\,27.\\
U.~M\"uller et al.:
{\sl The Omega RICH in the CERN hyperon beam experiment}.
{ Nucl.\ Instr.\ and Meth.\ }{\bf A433}\,(1999)\,71.

\bibitem{holroyd}
R.A.~Holroyd et al.:
{\sl Measurement of the absorption length and absolute quantum efficiency
of TMAE and TEA from threshold to $120\,\mbox{nm}$}. 
{ Nucl.\ Instr.\ and Meth.\ }{\bf A261}\,(1987)\,440.

\bibitem{propselex}
J.~Russ et al.: {\sl A proposal to construct SELEX}. { Fermilab~P781}\,(1987),
unpublished.\\
J.~Russ: {\sl Fermilab Hyperon Program: Present and Future Plans.}
{ Nucl.\ Phys.\ }{\bf A585}\,(1995)\,39c.


\bibitem{selexproto}
M.P.~Maia et al.: {\sl A Phototube RICH Detector}.
{ Nucl.\ Instr.\ and Meth.\ }{\bf A326}\,(1993)\,496.

\bibitem{sphinx}
V.A.~Dorofeev et al.:
{\sl The Search for heavy Pentaquark Exotic Baryons with hidden Strangeness 
in the $p + n \to (p \Phi) + n$ and 
$p + n \to (\Lambda(1520)\, K^+) + n$ Reactions at $E(p) = 70\,\mbox{GeV}$}.
{ Physics of Atomic Nuclei} {\bf 57}\,(1994)\,227.\\
A.~Kozhevnikov et al.:
{\sl SPHINX Phototube RICH Detector for Diffractive Production Experiments at
 Serpukhov Accelerator}.
{ Nucl.\ Instr.\ and Meth.\ }{\bf A433}\,(1999)\,164.

\bibitem{selexgas}
R.~Richardson and R.~Schmitt:
{ Adv.\ in Cryo.\ Eng.\ }{\bf 41B}\,(1996)\,1907.

\bibitem{ronchi}
L.~Stutte, J.~Engelfried and J.~Kilmer:
{\sl A Method to evaluate Mirrors for Cherenkov Counters}.
{ Nucl.\ Instr.\ and Meth.\ }{\bf A369}\,(1996)\,69.

\bibitem{partid} U.~M\"uller et al.:
{\sl Particle identification with the RICH detector in experiment
WA89 at CERN}.
{ Nucl.\ Instr.\ and Meth.\ }{\bf A343}\,(1994)\,279.

\bibitem{babar} I.~Adam et al.: 
{\sl Operation of the Cherenkov Detector DIRC of BABAR at
High Luminosity}. SLAC--Pub--8783 (2001).

\bibitem{schwiening}
J.~Schwiening: {\sl The DIRC Detector at the SLAC B-Factory PEP-II:
Operational Experience and Performance for Physics Application}
Nucl.\ Instr.\ and Meth.\ {\bf A502} (2003) 67.
Preprint SLAC-PUB-7473.


\bibitem{ckmprop}
CKM Collaboration, R.~Coleman et al.:
{\sl CKM -- Charged Kaons at the Main Injector --
A proposal for a Precision Measurement of the Decay
$K^+ \to \pi^+ \nu\bar\nu$ and
Other Rare $K^+$ Processes at Fermilab Using the Main Injector}.
{ FERMILAB-P-0905} (1998), unpublished.\\
CKM Collaboration, J.~Frank et al.,
{\sl CKM - Charged Kaons at the Main Injector - A proposal for a
Precision Measurement of the Decay 
$K^+ \to \pi^+ \nu\bar\nu$ and Other Rare $K^+$
Processes at Fermilab Using the Main Injector}.
Proposal (2nd edition) FERMILAB-P-0921, April 2001. 


\bibitem{ckmrich}
J.~Engelfried et al.:
{\sl Two RICH Detectors as Velocity Spectrometers in the CKM Experiment}.
Nucl.\ Instr.\ and Methods {\bf A502} (2003) 62-65.
hep-ex/0209020.


\end{thebibliography}
\end{document}